\documentclass[aps,preprint]{revtex4}%
\usepackage{amsfonts}
\usepackage{amsmath}
\usepackage{amssymb}
\usepackage{graphicx}%
\begin{document}

\begin{center}
\textbf{Scaling functional patterns of skeletal and cardiac muscles: New
non-linear elasticity approach}
\end{center}

.

\begin{center}
Valery B. Kokshenev

.

\textit{Departamento de Fisica, Instituto de Ciencias Exatas, Universidade
Federal de Minas Gerais, Caixa Postal 702, 30123-970, Belo Horizonte, Minas
Gerais, Brazil. Email: valery@fisica.ufmg.br}

.

Submitted to the Physica D 15 August 2009
\end{center}

.

\textbf{Abstract}: Responding mechanically to environmental requests, muscles
show a surprisingly large variety of functions. The studies of \textit{in
vivo} cycling muscles qualified skeletal muscles into four principal locomotor
patterns: motor, brake, strut, and spring. While much effort of has been done
in searching for muscle design patterns, no fundamental concepts underlying
empirically established patterns were revealed. In this interdisciplinary
study, continuum mechanics is applied to the problem of muscle structure in
relation to function. The ability of a powering muscle, treated as a
homogenous solid organ, tuned to efficient locomotion via the natural
frequency is illuminated through the non-linear elastic muscle moduli
controlled by contraction velocity. The exploration of the elastic force
patterns known in solid state physics incorporated in activated skeletal and
cardiac muscles via the mechanical similarity principle yields analytical
rationalization for locomotor muscle patterns. Besides the explanation of the
origin of muscle allometric exponents observed for muscles in legs of running
animals and wings of flying birds, the striated muscles are patterned through
primary and secondary activities expected to be useful in designing of
artificial muscles and modeling living and extinct animals.

.



\pagebreak

\section{Introduction}

The mechanical role of muscles varies widely with their architecture and
activation conditions. Striated (skeletal and cardiac) muscles are diverse in
their contractive interspecific and intraspecific functional properties
observed among and within animal species, nevertheless, "the smaller muscles
and muscles of smaller animals are quicker". This generic feature of skeletal
muscles was established by Hill [1]. More recently, the physiological
adaptation of skeletal muscles resulting in beneficial changes in muscle
function has been recognized by a number of investigators. It was learned that
long-fibre muscles commonly contract at over larger length ranges and
relatively higher velocities producing the greatest muscle forces the lowest
relative energetic costs [2]. Muscles having shorter fibres expose smaller
length change, but their cost of force generation is relatively less,
\textit{e.g.} [3]. Searching for determinants of evolution of shape, size, and
force output of cardiac and skeletal muscle, a little is known about the
regulation of directional processes of mass distribution [4,5]. Although
skeletal muscles grow in length as the bones grow, most studies only involve
force increasing with cross-sectional area. Following the idea that the muscle
force production function is a critical evolutionary determinant [5], I
develop a physical study of muscle form adaptation to a certain primary
activity with growth of size preserving muscle shape.

When designing architecture of the striated muscle built from repeating units
(fibres and sarcomeres) at least three distinct muscle activities should be
distinguished [5]: (i) the \emph{concentric }contraction defined as the
production of active tension while the muscle is shortening and performing
positive work, (ii) the \emph{eccentric }contraction defined as the
contraction during lengthening performing negative work in a controlled
fashion, and (iii) the \emph{isometric }contraction when the muscle force
output is produced without changing of length and performing net work. The
corresponding \emph{mechanical work patterns} called by Russel \textit{et al.}
[5] as "concentric work" and "eccentric work" (that might be extended by
"isometric work") were carefully studied via \textit{in vivo} measurements of
length-force cycling of individual skeletal muscles in active animals, such as
(i) the pectoralis in flying birds, (ii) leg extensors in running cockroaches,
and (iii) gastrocnemius in the level running turkey. The corresponding
\ \emph{muscle locomotor patterns} were called as (i) \emph{motor}, (ii)
\emph{brake}, and (iii) \emph{strut} functions [6].

The seminal research by Hill [1] on dynamics of electrically stimulated
\emph{isolated} muscles was restricted to a single isotonic shortening. The
studies of the corresponding motor function resulted in famous
force-inverse-velocity master curve remaining the major dynamic constraint of
all real (slow-fibre, fast-fibre, and superfast) muscles [7] and
computationally modeled muscles, \textit{e.g.} [8]. Besides, other two
fundamental rules of muscle dynamics were noted by Hill [1]. Examining
hovering humming and sparrow birds, he recognized that the "frequencies of
wings are roughly in inverse proportion to the cube roots of the weights,
\textit{i.e.} to linear size". Moreover, because the linear proportionality
between the stroke period and body length was found equally in electrically
stimulated isolated muscles, the intrinsic \emph{frequency-length} feature
constrained by scaling rule $T_{m}^{-1}\varpropto L_{m}^{-1}$ beyond the
nervous control is likely more universal than previously appreciated. Second
\emph{velocity-length} Hill's constraint states that "the intrinsic speed of
muscle has to vary inversely to length", \textit{i.e.} $V_{m}\varpropto
L_{m}^{-1}$. Both Hill's scaling rules remain a challenge to viscoelastic
models of transient-state mechanics and other existing theories of muscle
contraction [9].

The earliest theories of muscle motor function supposed muscle to be an
elastic body which, when stimulated, was converted in an active state
containing elastic energy causing the muscle to shorten. Such
\emph{elastic-energy theories} failed to explain mechanisms of the force
production in terms of viscoelastic characteristics. To a certain extent, poor
experimental approaches providing often conflicting clues to muscle structure
in relation to function may explain a little progress in understanding of
contractile properties of a muscle [9,4]. Moreover, physiological muscle
properties accounted for theories of muscle contractions developed at both
molecular and macroscopic scales are primarily focused on the reproduction of
force-velocity curve [9]. Besides, the existing phenomenological frameworks
such as Hill-type muscle models only mimic the proper mechanical
characteristics of muscles by means of passive viscoelastic springs attached
to muscle contractive element in series [10,11,3] or in parallel [12] and
recruited when muscle is activated. By ignoring the proper muscle function of
force production and force transmission throughout the muscle organ, these
models are able to explain no one of Hill's principal constraints in muscle
dynamics. On the other hand, there exist experimental evidences of the
adaptive ability of skeletal muscle to exchange \emph{elastic} strain energy
during force production [2]. In line with this concept, it has recently
communicated on a possibility of the modeling of the adaptive muscle
elasticity by elastic force patterns [13].

In the present paper, I develop an integrative theoretical framework to the
problem of forces, structure, and contractive non-linear dynamics in striated
muscles. Instead of Hill-type modeling of \textit{in vitro} motor function,
\textit{e.g.} [3], brake function, \textit{e.g.} [12, 2], and strut function,
\textit{e.g.} [14], or study of muscle design by means of simulation of
phenomenological force-length and/or force-velocity constraints [8], the
powerful method of continuum mechanics generally providing macroscopic
characterization and modeling of soft tissues, \textit{e.g.} [15, 16], is
employed. By further exploration of the elastic force patterns, I propose a
self-consistent depiction of the three velocity-distinct characteristic points
well distinguished in all \textit{in vivo} force-length loops of the naturally
activated muscles. Unlike the earliest elastic theories based on minimization
of energy, I develop the physical concept of similarity between the force
output and reaction active elastic forces that permits to avoid the details of
muscle activation process. The theory is validated by a comparison to
phenomenological scaling rules including both mentioned Hill's dynamic
constraints and therefore may be hopefully helpful in designing artificial
muscles [15] and modeling living and extinct organisms [17].

\section{Theory}

\subsection{Theoretical Background}

\subsubsection{McMahon's scaling models}

The engineering models by McMahon [18, 19] develop previous Hill's approach to
the problem of scaling of parameters of animal performance to \emph{body
weight} $W=Mg$. Using Hill's geometric similarity models [1,19] equally
applied to animal body, long bone, or individual muscle, each one was
approximated by a cylinder of longitudinal \emph{length} $L$ and
\emph{cross-sectional area} $A$ (or diameter $D\backsim\sqrt{A}$). Then, the
assumption on the \emph{weight-invariance} of for the \emph{tissue density},
namely%
\begin{equation}
\rho_{tiss}=\frac{M}{AL}\varpropto W^{0}\text{,} \label{Ro-m}%
\end{equation}
was adopted. In mammalian long-bone allometry, this invariant was verified and
observed with a high precision \ [20]. Mechanical models of bending bones and
shortening muscles were introduced by McMahon via the weight-invariant
\emph{elastic modulus }$E_{tiss}$, \emph{\ stress} $\sigma_{tiss}$, and
\emph{strain} $\varepsilon_{tiss}$, namely
\begin{equation}
E_{tiss}=\frac{\sigma_{tiss}}{\varepsilon_{tiss}}\varpropto W^{0}\text{, with
}\sigma_{tiss}=\frac{\Delta F}{A}\text{ and }\varepsilon_{tiss}=\frac{\Delta
L}{L}\text{.} \label{E-m}%
\end{equation}
Here $\Delta L$ ($=L-L_{0}$) is the \emph{length change }accompanied by the
\emph{force change}\ $\Delta F$ ($=F-F_{0}$) counted off from the
\emph{resting length} $L_{0}$.

While searching for functional mechanical patterns of biological systems
determined by \emph{maximal} forces using Eqs. (\ref{Ro-m}) and (\ref{E-m}),
the maximal stress/strain scaling relations%
\begin{equation}
\sigma_{geom}^{(\max)}\varpropto W^{1/3}\text{, }\sigma_{elast}^{(\max
)}\varpropto W^{1/4}\text{, and }\sigma_{stat}^{(\max)}\varpropto
W^{1/5}\text{,} \label{McMag-stress}%
\end{equation}
could be readily derived from McMahon's \emph{geometric }(isometric),
\emph{elastic} (buckling stress) and \emph{static} (bending elastic stress)
\emph{similarity models} distinguished through McMahon's scaling relations
\begin{equation}
L_{geom}\backsim D\text{, }L_{elast}\varpropto D^{2/3}\text{, and }%
L_{stat}\varpropto D^{1/2}\text{.} \label{Lmod}%
\end{equation}
Instead, the \emph{maximum} stress and strain
\begin{equation}
\sigma_{tiss}^{(\max)}\text{ }\varpropto\varepsilon_{tiss}^{(\max)}\varpropto
W^{0}\text{,} \label{P-const}%
\end{equation}
were postulated (see Table 4 in [19]) extending groundlessness his exact
result for the \emph{mean} stress $\sigma_{elast}^{(mean)}\varpropto W^{0}$,
obtained within the static stress similarity model (see Fig. 1 in [19]). The
improved self-consistent maximal stresses shown in Eqs. (\ref{McMag-stress})
follow straightforwardly from McMahon's cross-sectional areas%
\begin{equation}
A_{geom}^{(isom)}\varpropto W^{2/3}\text{, }A_{elast}^{(buck)}\varpropto
W^{3/4}\text{, and }A_{static}^{(bend)}\varpropto W^{4/5} \label{Amod}%
\end{equation}
applied to Eq. (\ref{E-m}), along with McMahon's idea on the dominating
gravitational forces in bones, muscles, and bodies, \textit{i.e.} $\Delta
F\backsim gM_{b}\backsim gM_{m}\backsim W$. As shown in [20], the structure of
long bones is driven by peak muscle forces, but not by gravity.

\subsubsection{Muscle shape and structure}

After Alexander [21], the \emph{physiologic} cross-sectional area $A_{0m}$
(PCSA) of the isolated skeletal \emph{muscle }$m$\emph{\ }of\emph{\ mass}
$M_{m}$ composed of $N$ bundles of masses $m_{i}$ was commonly estimated,
\textit{e.g.} [22], with the help of the cylinder-geometry relation
$A_{i}=m_{i}/\rho_{musc}L_{i}$, where $L_{i}$ \ is directly measured muscle
fibre length. The spindle-like shape of the muscle as whole organ was
therefore determined by the muscle PCSA, namely\emph{\ }
\begin{equation}
\text{ }A_{0m}=%
{\displaystyle\sum\limits_{i=1}^{N}}
A_{i}=\frac{M_{m}}{\rho_{musc}L_{0m}}\text{, and }\frac{1}{L_{0m}}=\frac
{1}{M_{m}}%
{\displaystyle\sum\limits_{i=1}^{N}}
\frac{m_{i}}{L_{i}}\text{, with }M_{m}=%
{\displaystyle\sum\limits_{i=1}^{N}}
m_{i}\text{,} \label{ro-m}%
\end{equation}
resulted in the sum of areas of muscle and the muscle length $L_{0m}$ of the
parallel-linked contractible subunits described statistically by the
length-unversed sum weighed by masses. Such a coarse-grained characterization
of the \emph{muscle structure }generally ignores the arrangement of muscle
fibres relative to generated force axis, distinguished by \emph{pinnate
angles}.

In scaling models, the evolution of the muscle structures across
different-sized animals of \emph{body mass} $M$ is observed statistically via
\emph{allometric exponents} $a_{m}$, $l_{m}$, and $\alpha_{m}$ determined by
common rules [21,23,25]:
\begin{equation}
A_{0m}\varpropto M_{m}^{a_{m}}\text{, }L_{0m}\varpropto M_{m}^{l_{m}}\text{,
and }M_{m}\varpropto M^{1+\alpha_{m}}, \label{am-lm}%
\end{equation}
where the \emph{muscle mass index} $\alpha_{m}$ plays the same role as
Prangel's index $\beta$ in bones, as noted in [26].

When the muscle-density invariance employed implicitly in Eq. (\ref{ro-m}) and
specified in Eq. (\ref{Ro-m}) is applied to different skeletal muscles, the
muscle shape approximated by cylinder geometry is also preserved.
Consequently, the \emph{muscle functional volume}%
\begin{equation}
A_{m}L_{m}=A_{0m}L_{0m}=\frac{M_{m}}{\rho_{0m}}\text{, with }\rho_{0m}%
=\rho_{musc}\varpropto M_{m}^{0}\varpropto M^{0}\text{,} \label{Mmu}%
\end{equation}
holding in all muscle work loops plays the role of the mechanical muscle
invariant. This condition is ensured by functional change $\Delta\rho
_{musc}/\rho_{musc}$ not exceeding $5\%$ [24]. Hence, the function-independent
\emph{muscle-shape constraint} [13]%
\begin{equation}
a_{m}+l_{m}=1+\alpha_{m} \label{am-lm-new}%
\end{equation}
straightforwardly follows from Eqs. (\ref{am-lm}) and (\ref{Mmu}). Likewise
the case of hindlimb mammalian bones of the mean structure $a_{b}^{(\exp
)}=2d_{b}^{(\exp)}=0.752$, $l_{b}^{(\exp)}=0.298$, and $\beta^{(\exp)}=0.04$
[20,26], Eq. (\ref{am-lm-new}) is also empirically observable in muscle
allometry (see analysis in Table 5 below).

\subsection{General Muscle Characterization}

\subsubsection{Maximal force and stress}

In \textit{in vivo} work loops, the muscle locomotor patterns can be generally
specified regardless of details of activation-deactivation conditions. In Fig.
1, the linear-slope characteristics $L_{1m}$ can be introduced in the
force-length cycling by the domains: $L_{2m}<L_{1m}<L_{3m}\approx L_{0m}$, for
the motor function, $L_{2m}>L_{1m}>L_{3m}$, for the brake function, and by
$L_{2m}\gtrsim L_{1m}\gtrsim L_{3m}\approx L_{0m}$, for the strut function
showing nearly isometric muscle contractions.

.

\textbf{Place Fig. 1}

.

Moreover, such a qualitative general characterization of the activated
individual muscle $m$ of \emph{resting length} $L_{0m}$ can be rationalized on
the basis of common \emph{two-point} force-length characterization, namely%
\begin{equation}
F_{musc}^{(\exp)}(L_{2m})=F_{musc}^{(\max)}=F_{2m}\text{ and }F_{musc}%
^{(\exp)}(L_{1m})=F_{1m}\text{,} \label{Lopt}%
\end{equation}
\emph{\ }introduced by the maximum force\emph{\ }$F_{2m}^{(\max)}$ and the
\emph{optimum }muscle length [28, 29] $L_{1m}$ . The \emph{instant dynamic
length} $L_{m}=L_{1m}\pm\Delta L_{1m}$ is counted off from the characteristic
point $L_{1m}$ via the optimum length change $\Delta L_{1m}$ (\ref{Lopt})
shown in Fig. 1 for all functions.

First, the linearization of the \textit{in vivo} muscle force-length curve
allows one to determine \emph{trial} peak stress and corresponding strain by%
\begin{equation}
\sigma_{musc}^{(\max)}=\frac{F_{musc}^{(\max)}}{A_{2m}}\text{ and }%
\varepsilon_{musc}^{(\max)}=\frac{\Delta L_{1m}^{(\max)}}{L_{2m}}\text{, with
}\Delta L_{1m}^{(\max)}=|L_{2m}-L_{1m}|\text{. } \label{Sm}%
\end{equation}
The corresponding force change $\Delta F_{musc}^{(\max)}$ observed near the
\emph{optimum force} $F_{1m}^{(\max)}$ (\ref{Lopt}) provides%
\begin{equation}
F_{musc}^{(\max)}=F_{musc}^{(\exp)}(L_{1m})+\Delta F_{musc}^{(\max)}%
=F_{1m}+K_{musc}^{(\max)}\Delta L_{1m}^{(\max)}\text{ } \label{Fmax}%
\end{equation}
that determinates \emph{effective} \emph{muscle stiffness }and \emph{effective
modulus}, respectively%
\begin{align}
K_{musc}^{(\max)}  &  \equiv K_{2m}=\left\vert \frac{dF_{musc}}{dL_{m}%
}\right\vert _{F_{1m}^{(\max)}}\thickapprox\frac{\Delta F_{musc}^{(\max)}%
}{\Delta L_{1m}^{(\max)}}=\frac{\Delta F_{musc}^{(\max)}}{F_{musc}^{(\max)}%
}E_{musc}^{(\max)}\frac{A_{2m}}{L_{2m}}\text{, }\nonumber\\
\text{and }E_{musc}^{(\max)}  &  \equiv E_{2m}=\frac{\sigma_{musc}^{(\max)}%
}{\varepsilon_{musc}^{(\max)}}\text{,} \label{k-musc}%
\end{align}
following from Eqs. (\ref{Sm}) and (\ref{Fmax}).

\subsubsection{Active stiffness and resonant muscle mechanics}

Secondly, treating the maximum-force crossover state (\ref{Lopt}) as the
generic transient-neutral\emph{\ }state [26] the \emph{resonant}
\emph{frequency} $1/T_{musc}^{(\max)}=T_{2m}^{-1}$ related to point $2$ in
Fig. 1 associated with maximum efficiency of muscle cycling, \textit{e.g.}
[29], can also be introduced as \emph{natural frequency} [19], namely%
\begin{equation}
T_{2m}^{-1}\thicksim2\pi\sqrt{\frac{K_{2m}}{M_{m}}}\thicksim\sqrt
{\frac{E_{musc}^{(\max)}}{\rho_{0m}}}\left(  \frac{\Delta F_{musc}^{(\max)}%
}{F_{musc}^{(\max)}}\right)  ^{1/2}\frac{1}{L_{2m}}\text{,} \label{Freq}%
\end{equation}
and analyzed by Eqs. (\ref{Mmu}) and (\ref{k-musc}).

One can see that Eq. (\ref{Freq}) yields first Hill's general constraint
discussed in Introduction. However, the following three conditions are
required: (i) the preservation of dynamic functional volume\emph{\ }%
(\ref{Mmu}), (ii) the weight-invariance of the elastic modulus $E_{musc}%
^{(\max)}$ (\ref{E-m}), and (iii)\ the existence of force similarity between
the exerted force $F_{musc}^{(\max)}$ and its change $\Delta F_{musc}^{(\max
)}$ (\ref{Fmax}). Therefore, the muscle \emph{force-similarity principle},
namely%
\begin{equation}
F_{musc}\cong\Delta F_{musc}\cong F_{prod}\cong F_{elast}\cong\Delta
F_{elast}\text{,} \label{Fm-elast}%
\end{equation}
implying a coexistence of all forces in biomechanically equivalent states [26]
must be adopted. Here the \emph{active elastic force} $\Delta F_{elast}$
(shown schematically as $F_{act}$ in Fig. 1D) is also included. The total
state-transient elastic force $F_{elast}$ is the superposition of common
\emph{passive elastic force} $F_{pass}$\ provoked by external loads and
\emph{active elastic force} $\Delta F_{elast}$\ caused by the
\emph{production} force $F_{prod}$. The correspondence sign $\cong$ indicates
that though the involved physical characteristics belong to the same
mechanical state, they may differ in both physical and numerical parameters
stipulating this state.

Given that the peak \emph{active muscle stress} $\sigma_{m}$ always exceeds
the corresponding passive stress, \textit{e.g.} [14], in further I focus on
the fully activated \emph{transient states} described by
\begin{equation}
\sigma_{m}=\frac{\Delta F_{elast}}{A_{m}}=E_{m}\frac{\Delta L_{m}}{L_{m}%
}\text{. } \label{sigma-el}%
\end{equation}
Unlike Eqs. (\ref{E-m}) and (\ref{Sm}), $\sigma_{m}$\ is the true intrinsic
\emph{elastic stress} in a certain (not specified) dynamic state. This reveals
the maximum-amplitude \emph{elastic force\ }of the fully activated muscle%
\begin{equation}
\Delta F_{elast}\equiv\Delta F_{m}=K_{m}\Delta L_{m}=E_{m}A_{m}\frac{\Delta
L_{m}}{L_{m}} \label{Felast}%
\end{equation}
and in turn provides the corresponding \emph{active} \emph{muscle stiffness}%
\begin{equation}
K_{m}=E_{m}\frac{A_{m}}{L_{m}}\text{.} \label{Kopt}%
\end{equation}
The underlying mechanical \emph{sarcomere elastic stiffness} $K_{s}$ is
related via the muscle-volume average, namely
\begin{equation}
K_{m}=\frac{1}{A_{m}L_{m}}\int K_{s}(r_{m})\text{ }d^{3}r_{m}\text{,}
\label{Km-Ks}%
\end{equation}
originated from end-to-end intercellular overlapping [31, 12].

The \emph{muscle energy change}%
\begin{equation}
\Delta U_{m}\backsim K_{m}\Delta L_{m}^{2}\cong E_{m}A_{m}\frac{\Delta
L_{m}^{2}}{L_{m}} \label{U-m}%
\end{equation}
stored or released during active-period contraction provides the mechanical
\emph{cost of energy}%

\begin{equation}
CU_{m}=\frac{\Delta U_{m}}{\Delta L_{m}}\cong E_{m}A_{m}\frac{\Delta L_{m}%
}{L_{m}}\text{.} \label{CUm}%
\end{equation}
These relations demonstrate how the observable mechanical characteristics can
be linked to the underlying muscle elastic forces using the force-similarity
principle (\ref{Fm-elast}). In turn, the \emph{contraction velocity}%
\begin{equation}
V_{m}=\overline{V_{m}(t)}\equiv\frac{1}{\Delta t_{m}}\int_{0}^{\Delta t_{m}%
}\left[  \frac{dL_{m}(t)}{dt}\right]  dt\backsim\left[  \frac{dL_{m}(t)}%
{dt}\right]  _{t=\Delta t_{m}}\cong\frac{L_{m}}{T_{m}} \label{Vcont}%
\end{equation}
is defined by the instant velocity $V_{m}(t)$ averaged over \emph{activation
time} $\Delta t_{m}$.

\subsubsection{Fast and slow activated muscles}

According to the most general classification of diverse muscles, three types
are conventionally distinguished: red (slow fibre) muscles, white (fast fibre)
muscles, and intermediate type, mixed fibre muscles. Although collective
mechanisms of muscle contractions are poor understood, \textit{e.g.} [32],
physically, the two limiting situations of dynamic accommodation of local
forces generated by cross bridge attachments can be generally rationalized. As
schematically drawn in Fig. 1D, in an activated muscle, the dynamic process of
equilibration between the production intrinsic forces and external loads (not
shown) is followed by the spatiotemporal relaxation of elastic forces. For the
simplest case of \emph{slow muscles}, the dynamic equilibration occurs via the
slow channel of relaxation, assumably common for both active, $F_{prod}%
^{(slow)}$, and passive elastic forces. Since passive forces in solids are
short of range [33], both the forces are proportional to muscle \emph{surface}%
. In contrast, it is plausible to adopt that in \emph{fast muscles} the
fast-twitch fibres transmit the locally generated forces in all directions,
\textit{i.e.} along and across fibres, resulting in the overall maximum force
output $F_{prod}^{(fast)}$ to be linear with dynamic muscle \emph{volume}.
Basing on such a general physical picture, a function-independent and regime-
independent characterization of the force production function, namely
\begin{equation}
F_{prod}^{(fast)}\varpropto A_{rm}L_{rm}\text{ and }F_{prod}^{(slow)}%
\varpropto A_{rm}\text{, with }r=1,2\text{, and }3\text{,} \label{Ffaslo}%
\end{equation}
is proposed via the force-size scaling rules for all three distinct states
shown in Fig. 1 and hereafter distinguished by symbol $r$.

The widely adopted by biologists linear-displacement regime is discussed in
Eq. (\ref{E-m}) via $\Delta L\varpropto L$ resulted in the weight-independent
strain (\ref{P-const}). The corresponding optimum-velocity regime $r=1$,
attributed to the instant length-independent elastic strains, $\varepsilon
_{m}^{(opt)}=|L_{m}-L_{3m}|/L_{m}\varpropto L_{m}^{0}$ (\ref{E-m}) with
$L_{m}$\ lying between\textrm{\ }$L_{1m}$\textrm{\ }and $L_{3m}\approx L_{0m}%
$, is now clarified by the scaling equations%
\begin{equation}
E_{1m}^{(fast)}=E_{fast}^{(opt)}\varpropto L_{m}^{1}\text{ and }%
E_{1m}^{(slow)}=E_{slow}^{(opt)}\varpropto L_{m}^{0} \label{E1}%
\end{equation}
characteristic of fast and slow muscles. Such a muscle description follows
from the similarity (\ref{Fm-elast}) between the active elastic force $\Delta
F_{1m}=\Delta F_{elast}^{(opt)}=E_{1m}A_{1m}\varepsilon_{1m}$ (\ref{Felast})
and corresponding production force (\ref{Ffaslo}). The \emph{optimum}
force-velocity muscle mechanics is rationalized below in\textbf{ }Table 1 and
then tested by empirical data.

Similarly, the bilinear-displacement regime $r=2$ introduced by the dynamic
length change $\Delta L_{2m}=|L_{m}-L_{1m}|\varpropto L_{m}^{2}$, with $L_{m}
$\ lying between $L_{2m}$\ and $L_{1m}$, and the \emph{maximum} active elastic
force $\Delta F_{2m}=\Delta F_{elast}^{(\max)}=E_{2m}A_{2m}\varepsilon
_{2m}^{(\max)}$ (\ref{Felast}) results in the maximal elastic moduli%
\begin{equation}
E_{fast}^{(\max)}=E_{2m}^{(fast)}\varpropto L_{m}^{0}\text{ and }%
E_{slow}^{(\max)}=E_{2m}^{(slow)}\varpropto L_{m}^{-1}\text{,} \label{E2}%
\end{equation}
adjusted with the muscle production function (\ref{Ffaslo}) via the force
similarity principle (\ref{Fm-elast}). Finally, the high-velocity trilinear
regime $r=3$ is suggested by the moderate-force and moderate-elastic muscle
determined by%
\begin{equation}
E_{fast}^{(\operatorname{mod})}=E_{3m}^{(fast)}\varpropto L_{m}^{-1}\text{ and
}E_{slow}^{(\operatorname{mod})}=E_{3m}^{(slow)}\varpropto L_{m}^{-2}\text{. }
\label{E3}%
\end{equation}
This condition specifies point $3$ in Fig. 1, along with the underlying
cubic-power muscle displacements $\Delta L_{3m}\varpropto L_{m}^{3}$ scaled by
dynamic $L_{m}$ lying\textrm{\ }above or below the characteristic length
$L_{3m}$\ in any muscle acting as motor, brake or strut (see Fig. 1).

\subsection{Muscle Functions}

Likewise the naturally curved mammalian long bones biomechanically adapted to
the maximum longitudinally bending [20, 26], the muscle\emph{\ motor function}
is assigned to locomotor muscles showing concentric positive work exerted by
elastic bending forces. Given that the \emph{elastic force patterns} coincide
for bending and torsion [26], both kinds of unpinnate and uni-pinnate skeletal
muscles, having respectively close to zero and non-zero fixed pinnate angles,
may be expected to be structured by the same motor function. The
specific-function mechanical characterization is described in Appendix B and
results are summarized in Table 2.

\section{Results}

\subsection{Assumptions and predictions}

The following assumptions are made regarding elastic striated muscles:

1. The muscles are considered at macroscopic scale as individual homogeneous
organs. Within the continuum mechanics, the coarse-grained approach ignores
the details of heterogeneous microstructure and pinnate angles.

2. When activated under different boundary loaded conditions, the muscles do
not undergo changes in shape and whole volume. The emerging elastic fields
follow patterns established for long solid cylinders.

3. The mechanical similarity adopted between the extrinsic forces exerted by
the muscle and intrinsic elastic reaction forces, as well as the dynamic
similarity adopted for contraction velocities and frequencies are observable
in all biomechanically equivalent states.

4. The natural ability of the non-linear elastic tuning of fast and slow
muscles to distinct locomotor states can be characterized by the elastic
moduli sensitive to evolving dynamic variable\ associated here with the
regime-characteristic muscle length.

.

The function-independent mechanical characterization of muscles is provided in
Table 1.

.

\textbf{Place Table 1}\textrm{. }

\textrm{.}

The specific case of muscle structure accommodation in the bilinear regime is
described in Table 2.

.

\textbf{Place Table 2}

.

The rules of mass distribution\textrm{\ }across and along the muscle axis
provided in Table 2 in terms of the muscle-structure scaling exponents
[$a_{2m}$, $l_{2m}$] are characteristic for slow, fast, and mixed muscles
producing maximum force. In Table 3, these scaling rules are compared with the
finding for the optimal-force state [$a_{1m}$, $l_{1m}$] and moderate-force
state [$a_{3m}$, $l_{3m}$].

.

\textbf{Place Table 3}

\textrm{.}

The dynamic characteristics of distinct-velocity contractions are predicted in
Table 4.

.

\textbf{Place Table 4}

.

The consequences of the theoretical scaling framework are:

1. The peak forces generated in all regimes scales as muscle volume or surface
in fast or slow muscles, respectively.

2. A general, function-independent mechanical description of the striated
muscle activated in the liner-displacement regime is predicted for each type
of muscles (Table 1).

3. The muscle-type independent locomotor functions and related mechanical and
dynamic characteristics of the striated muscle activated in the bilinear
regime are predicted (Table 2).

4. The muscle-type independent varied dynamic structures are predicted for all
muscle regimes and functions (Table 3).

5. The function-independent dynamic scaling characteristics are obtained in
Table 4 for all type of muscles.

In what follows, all theoretical findings are tested by the available from the
literature data.

\section{Discussion}

"What determines the shape, size, and force output of cardiac and skeletal
muscle?" (Louis Sullivan quoted in [5]). The provided coarse-grained study of
conservative striated muscles suggests that the size-dependent peak elastic
forces determine fiber-type-independent patterns of the functionally adapted
structures preserving muscle shape. The size-dependent peak force output is
determined by the muscle volume and area for white and red muscles, regardless
of muscle structure and function.

\subsection{Function against structure}

\subsubsection{General muscle characterization}

Being composed of bundles of muscle fibres including all other contractible
components (neural, vascular, and collagenous reticulum), the striated muscle
is thought of as a heterogeneous\emph{\ continuum medium} transmitting the
produced tension internally and externally, \textit{e.g.} [34]. Primarily, I
address the problem of mechanical design of striated muscle to a general,
function-independent characterization of the individual muscle organ loaded by
tension, reaction, and gravity through tendons, ligaments, and bones. My
non-energetic approach is physically grounded by the existence of linear
force-displacement regions (shown by the solid arrows in Figs. 1A, 1B, and 1C)
in all \textit{in vivo} work loops regardless of dynamic details of
approaching to the maximum exerted force $F_{musc}^{(\max)}$. Hence, the
mechanical characterization of the maximum-force activated muscle arises from
the muscle\emph{\ }stiffness $K_{m}^{(\max)}$ (\ref{k-musc}) underlaid by
sarcomere stiffness $K_{s}^{(\max)}$ (\ref{Km-Ks}).\ Consequently, all forces
involved in muscle contraction following by active and passive elastic strains
allow common mechanical description (shown in Fig. 1D) not depending on their
biochemical, inertial, or reaction origin.

The analytical justification of Hill's first frequency-length constraint
arises from the analysis of Eq. (\ref{Freq}) that requires eventually the
usage of the similarity between all intrinsic muscle forces, Eq.
(\ref{Fm-elast}). The constraint $T_{m}^{-1}\varpropto L_{m}^{-1}$ and other
mechanical characteristics for slow muscles accumulated in Table can be
applied to \emph{steady-speed} regimes of locomotion modes where all forces
are generally equilibrated and controlled by slow-fibre muscles [35]. In the
case of non-steady transient locomotion when fast-twitch fibres and nervous
control are additionally requested [35], Hill's first constraint transforms
[by Eqs. (\ref{Freq}) and (\ref{E1})] into a new one, $T_{m}^{-1}\varpropto
L_{m}^{-1/2}\varpropto1/V_{fast}^{(opt)}$ (Tables 1 and 4), well known for
animals running with maximal \emph{optimum} speed [36,37] $V_{run}^{(\max
)}\backsim V_{fast}^{(opt)}\varpropto\sqrt{L}\varpropto\sqrt{L_{m}}$. We have
therefore demonstrated how the suggested \emph{dynamic similarity} establishes
a link between the body-propulsion speed and locomotor-muscle contraction
velocity, also described by Rome \textit{et al.} [38]. Being united with the
muscle-force similarity, both constraints yield \emph{mechanical similarity},
the key principle explored in this research.

\subsubsection{Maximum force output against structure and velocity}

In muscle physiology, the functional effect of muscle conceptual architecture
simply states that muscle force output is proportional to PCSA. The proposed
study of adaptation of the muscle structure via the force production function
seems to be in qualitative agreement with this statement, because in all cases
exposed in Eq. (\ref{Ffaslo}) the muscle force output is \emph{proportional}
to $A_{m}$. Such a simplified treatment of the fast-muscle mechanics (formally
substituted by that for slow muscles) arrived at the widely adopted opinion
that the peak muscle stress $F_{prod}^{(slow)}/A_{m}$ , specifying the case of
slow muscles in linear dynamic regimes with $\sigma_{m}^{(slow)}\varpropto
L_{m}^{0}$ (Table 1), is generic for any muscle, as already discussed in Eq.
(\ref{P-const}). Although the proposal on scaling of the maximum production
force (and active stress) with muscle size (\ref{Ffaslo}) is a challenge for
further research, the provided fairly general physical grounds are supported
by empirical observations by Marden and Allen [39]. They established
statistically that the maximum force output in all biological (and human-made)
motors falls into two fundamental scaling laws: (i) in fast-cycling motors,
presented by flying insects, bats and birds, swimming fishes, and\ running
animals it scales as (\emph{motor mass)}$^{1}$ and (ii) in slow-cycling
motors, such as myosin molecules, muscle cells, and some (unspecified)\ "whole
muscles" the force at output scales as (\emph{motor mass})$^{2/3}$. The "motor
mass" was associated with muscle (and fuel) mass. That fact that the authors
observed muscle motors from sarcomere to whole muscle organ passing through
the single-fibre level of muscle organization, makes a basis for the discussed
below \emph{micro-macro scale correspondence}.

The proposed treatment of the \textit{in vivo} force-length curves is provided
for three distinct force-velocity characteristic points (shown in Fig. 1)
correlated by the inequalities%
\begin{equation}
F_{2m}>F_{1m}>F_{3m}\text{ and }V_{2m}<V_{1m}<V_{3m}\text{.} \label{F1,2,3}%
\end{equation}
These three generic function-independent\ states are associated with the
linear ($r=1$), bilinear ($r=2$), and trilinear ($r=3$) muscle dynamics
determined via the muscle elastic moduli $E_{rm}$ in Eqs. (\ref{E1}),
(\ref{E2}), and (\ref{E3}), respectively. The mechanical characterization of
slow and fast striated muscles is therefore provided in terms of the maximum
($\Delta F_{2m}$), optimum ($\Delta F_{1m}$) and moderate ($\Delta F_{3m}$)
active elastic forces developed at the measurable maximum ($V_{3m}$), optimum
($V_{1m}$), and moderate ($V_{2m}$) contraction velocities (Table 4). The
stabilization of the dynamic regimes is expected at the natural frequencies,
which also are scaled in Table 4 to the dynamic length $L_{rm}$.

\subsubsection{Muscle functions against size and shape}

Searching for answer on "what features make a muscular system well-adapted to
a specific function?" [28], it has been shown preliminary [13] that such
features are related to natural conditions of the stabilization or tuning to
the moderate-velocity regime $r=2$ via the mean dynamic length of the
fast-twitch fibers adapted by the best way to one of the patterns of muscle
locomotor functions. In this study such features specify the role of
slow-twitch fibers.

The elastic-force patterns underlying concentric, eccentric, isometric, and
cardiac contractions are suggested in Eqs. (\ref{F-bend}), (\ref{Fm-stret}),
(\ref{Fm-strut}), and (\ref{Funk}), respectively. The solutions to the
muscle-force and muscle-shape constraints are accumulated\textrm{\ }in Table 2
as patterned functions well distinguished by the muscle \emph{structure
parameter} ($\eta_{m}=d\ln A_{m}/d\ln L_{m}$) established for the motor
($\eta_{1}=4$), brake ($\eta_{2}=3$), strut ($\eta_{3}=\infty$) skeletal
muscles, an extended by the pump ($\eta_{5}=1$) cardiac muscle and one spring
($\eta_{4}=2$) striated muscle. These structurally adapted muscles are thought
of as to be suited to efficient work during powering when, respectively,
shortening ($m=1$), lengthening ($m=2$), or remaining in the nearly isometric
dynamic state ($m=3$), high-pressure-resistant state ($m=4$), and likely
energy-saving state ($m=5$). The found new pump function is in accord with the
observation by Russel \textit{et al.} [5] that "the heart chamber, unlike
skeletal muscles, can extend in both longitudinal and transverse directions,
and cardiac cells can grow in length and width", that implies $\eta_{5}<$
$\eta_{1},\eta_{2}$, or $\eta_{3}$. Given that only a few patterns exist in
elastic theory of solids [26], it is not striking that the spring, brake, and
motor functions resembles McMahon's\ "geometric", "elastic", and "static"
stress similarities discussed in Eqs. (\ref{McMag-stress}) and (\ref{Lmod}).

In Table 3, conceivable stable\ dynamic structures corresponding to muscle
activity in different dynamic regimes are analyzed. As in the case of Table 2,
the solutions of dynamic constraints follow from the similarity between force
output (\ref{Fm-elast}) and elastic-force patterns. The resulting
\emph{dynamic }states are discussed in terms of the scaling exponents for the
muscle \emph{dynamic structure} [$A_{rm}$, $L_{rm}$] preserving muscle shape
and volume (\ref{Mmu}). Other related observable mechanical characteristics
are exemplified in Tables 1 and 2. The major outcome of the analysis in Tables
2 and 3 is that both slow-twitch and fast-twitch fibres belonging to the same
muscle $m$ should manifest concerted behavior coordinated by the dynamic
active elastic forces.

Another significant feature of the analysis in Table 3 is a striking
prediction of the mechanical functions which are expected to be shown by a
given striated muscle $m$ of certain specialization (\emph{primary functions}
indicated by regime $r=2$, see proof below) when its cycling dynamics is
switched to regimes $r=1$ and $3$ by tuning to the corresponding natural
frequencies $T_{rm}^{-1}$. In case of regime $r=1$, both types of arbitrary
slow muscle tuned to $T_{1slow}^{-1}$ and fast muscle tuned to $T_{1fast}%
^{-1}$ (Table 4) are expected to show maximum workloop efficiency when acting
as controlled spring. In the efficient nonlinear regime $r=3$ the slow and
fast struts ($m=3$ in Table 2) will not show another function, but any type of
brakes ($m=2$ in Table 2) will work as motors, whereas motor are expected to
expose a new function, say, $m=6$ [determined by $\eta_{6}=(6/7)/(1/7)=6$]
that is closer to the brake activity ($\eta_{2}=3$) than the strut ($\eta
_{3}=\infty$). The cardiac muscles seem to display a crucial dynamic state,
say $m=0$ with $\eta_{0}=0$, which flatters the heart. Such predicted
\emph{secondary functions} and unusual ($m=0$ and $6$) muscles adapted to new
functions is a challenge deserving further study by experimentalists.

\subsection{Direct observation of muscle specialization}

"If a muscle is specialized for a particular mechanical role how this is
reflected in it architecture?" [40]. The stated problem is approached here by
the comparative analysis between the muscle allometric exponents and those
predicted for particular efficient activities describing the trends of biomass
accommodation via PCSA and along a muscle.

\subsubsection{Isolated muscles in hindlimb of mammals and birds}

In Table 5, the \emph{morphometric data} on the allometric exponents for the
mean cross-sectional area $A_{0m}^{(\exp)}$ and length $L_{0m}^{(\exp)}$ of
four\ skeletal muscles in the mammalian hindlimb for $35$ quadrupedal species
of body-mass domain exceeding four orders in magnitude are studied.

.

\textbf{Place Table 5}

.

First, let us verify the cylinder-shape similarity of skeletal muscles
described by Eq. (\ref{Mmu}). The muscle mass index $\alpha_{0m}$ estimated in
Eq. (\ref{am-lm-new}) via experimental data $a_{0m}^{(\exp)}$ and
$l_{0m}^{(\exp)}$ is compared in Table 5 with the measured indexes
$\alpha_{0m}^{(\exp)}$.

.

\textbf{Place Fig. 2}

.

In Figs. 2 and 3, the method of determination of the primary mechanical
function is illustrated: the adapted muscle structure is indicated by the
appropriate theoretical point located most closely to the datapoint.

.

\textbf{Place Fig. 3}

.

The found reliable estimates $\alpha_{0m}^{(est)}$ were used then in the
muscle-function analysis in Figs. 2 and 3. The established small indices
$\alpha_{0m}$ generally validate the muscle biomechanics by proving a
high-precision observation of locomotory muscle patterns via muscle
morphometry and \emph{functional physiology}. This implies that the effect of
biomechanical adaptation of muscle design to active elastic forces
predominates over effects of biological adaptation assigned to small
$\alpha_{0m}^{(\exp)}$.

Secondly, the analysis in Figs. 2 and 3 indicates strong correlations between
the morphometrically characterized structure of skeletal muscle and one of the
primary locomotor functions described in Table 2. The primary functions
indicated in Table 5 are found with a high degree of certainty. Indeed, as
illustrated in Fig. 2, the deviations of distances measured along the dashed
line, corresponding to a given muscle, between the datapoint and distant
challengers for the primary function, from the smallest distance indicating
the primary candidate, always exceed the experimental uncertainty.

Thirdly, the found muscle mechanical specifications do not conflict with the
\emph{physiological categorization} established for joint extensors and
flexors, which muscle structures are shown to be adapted to the brake and
motor functions via activation of eccentric and concentric elastic forces. The
found structure parameter $\eta_{plant}\thickapprox18$ indicates the foot
support activity for plantaris as the primary function (Table 5) that is in
accord with \textit{in vivo} workloop presented in Fig. 1C. As shown in Table
3, the struts are most conservative muscles no changing their support function
in non-linear regimes. In contrast, the gastrocnemius in mammals manifests
their motor, strut, and brake functions in, respectively, uphill, level, and
incline running of animals. Through the motor adapted structure with
$\eta_{gast}\thickapprox\eta_{1}=4$, the analysis in Fig. 3 establishes the
motor activity for gastrocnemius as the primary function naturally selected
for the significant mechanical task of uphill running exploring the bilinear
muscle dynamics. The effective trilinear gastrocnemius-displacement dynamics
is most close to the brake-like activity $\eta_{6}=6$, attributed to the
secondary function of the motor experimentally observed in gastrocnemius of
incline running turkey [27] and hopping tammar wallabies [24].

In Fig. 4, the overall muscle peak stress data measured in limb muscles of
animals in strenuous activity, reviewed by Biewener [25], are re-examined and
re-analyzed accounting for the primary functions of hindlimb muscles
established in Table 5.

.

\textbf{Place Fig. 4}

.

The uphill-motor specialization of gastrocnemius is independently supported by
the compressive-stress analysis made in Fig. 4 for fast running, jumping, and
hopping mammals. The stress scaling exponents ($s_{m}$) predicted for the
motor ($s_{1}=1/5$), strut ($s_{3}=0$), and control ($s_{4}=0$) functions are
shown to be distinguishable in work-specific mammalian muscles described in
Table 2. Hence, although the overall-function data by Biewener [25] indeed
expose almost weight-independent muscle stress, earlier postulated by McMahon
in Eq. (\ref{P-const}) and only in part justified here by the slow-fibre
muscles (Table 1) and strut muscles (Table 2), the analyses in Fig. 4
demonstrates how the function-specific muscle stress may serve as a new tool
for the direct observation of muscle specialization ignored in all previous
overall-function analyses.

I have also investigated an interesting question: whether the primary function
established for a certain leg muscle in mammals specialized to fast running
coincides with that for the same muscle in birds? The pioneering data on
individual leg muscles in $8$ running birds, ranging in size from $0.1$
\emph{kg }quail to $40$ \emph{kg} ostrich, are analyzed in Table 6\ and Fig. 5.

.

\textbf{Place Table 6}

.

\textbf{Place Fig. 5}

.

In running and non-running birds (Fig. 5), the \emph{gastrocnemius} is
employed as the brake and spring, in contrast to the motor function in mammals
(Table 5). This is in accord with Bennett [23], who noted that "the full
force-generated capacity of gastrocnemius is only used occasionally, such as
during take-off, when a bird attempts to throw itself into the air". This
explains our indirect observation: the primary function of the gastrocnemius
in running specialists is attributed to the foot flexor in mammals and ankle
extensor in birds (Table 6). In \emph{non-running }birds, the legs are
designed to control the ground locomotion (Fig. 5), whereas the wings may
share motor and brake functions (Table 3), in accord with the review by
Dickinson \textit{et al.} [6].

\subsubsection{Micro-macro scale correspondence}

There are many striking examples when skeletal muscles\ expose adaptation to a
specific function, \textit{e.g.} [43, 3]. The striated muscles anatomically
suited to concentric or eccentric work [2] are structurally distinct having,
respectively, long thin cells or short wide cells [5]. This observation
suggests the \emph{microscopic level} of muscle-cell adaptation introduced
here by
\begin{equation}
A_{cell}^{(conc)}>A_{cell}^{(ecent)}\text{ and }L_{cell}^{(ecent)}%
>L_{cell}^{(conc)} \label{str-corr}%
\end{equation}
for the \emph{cellular} cross-sectional area $A_{cell}$ ($\equiv A_{s}$) and
length $L_{cell}$ ($\equiv L_{s}$). Adopting these function specific trends,
one may expect to observe the cell-structure parameters $\eta_{s}=4$ and $3$
for sarcomeres accommodated to efficient shortening or stretching of muscle as
a whole.

A general question arises whether allometric coefficients of proportionality
omitted above in all structure-function power-law (scaling) relations are also
attributed to active elastic strains accompanying maximum force production?
Or, alternatively, other microscopically justified mechanisms, \textit{c.f.}
[44], or additional parameters (such as pinnate angle) may result in different
general macroscopic consequences? Given the highly conservative nature of
contractive units of \emph{skeletal} muscles\ and their well pronounced
organization [25], the \emph{specific-function trends} of the muscle
cross-sectional area%
\begin{equation}
A_{strut}^{(\text{\textit{isom}})}>A_{motor}^{(conc)}>A_{brake}^{(eccen)}%
>A_{contr}^{(sprin)}\text{ } \label{A-pred}%
\end{equation}
and muscle-fibre length%
\begin{equation}
L_{contr}^{(sprin)}>L_{brake}^{(eccen)}>L_{motor}^{(conc)}>L_{strut}%
^{(\text{\textit{isom}})} \label{L-pred}%
\end{equation}
are generally expected from Table 2. The suggested trends become observable
via the primary functions established in Table 5 for gastrocnemius ($m=1$),
DDF ($m=1$), CDE ($m=2$), and plantaris ($m=3$), when the regression data [22]
on passive-muscle structure [$A_{0m}^{(\exp)}(M)$, $L_{0m}^{(\exp)}(M)$] are
taken additionally into consideration: $A_{plant}^{(\exp)}>A_{gast}^{(\exp
)}\gtrsim A_{DDF}^{(\exp)}>A_{CDE}^{(\exp)}$ and $L_{CDE}^{(\exp)}%
>L_{gast}^{(\exp)}\gtrsim L_{DDF}^{(\exp)}>L_{plant}^{(\exp)}$, starting with
$M>1$ $kg$.

Similarly, the trend for active stiffness%
\begin{equation}
K_{strut}^{(\max)}>K_{motor}^{(\max)}>K_{brake}^{(\max)}\text{ and, generally,
}K_{fast}^{(\max)}>K_{slow}^{(\max)}\text{ } \label{K-pred}%
\end{equation}
straightforwardly follows from Table 2. Given that the \emph{optimum velocity}
for fast fibres $V_{1m}\varpropto L_{m}^{1/2}$ (Table 1), Eq. (\ref{L-pred})
provides%
\begin{equation}
V_{brake}^{(opt)}>V_{motor}^{(opt)}>V_{strut}^{(\text{\textit{opt}})}
\label{V-pred}%
\end{equation}
Moreover, a crude estimate for the \emph{cost energy}%
\begin{equation}
CU_{motor}^{(\max)}>CU_{strut}^{(\max)}>CU_{brake}^{(\max)} \label{CU-pred}%
\end{equation}
follows from $CU_{fast}^{(\max)}\varpropto M_{m}$ (\ref{CUm}) and the
experimental data by Pollock and Shadwick [22], $M_{1}^{(\exp)}>M_{3}^{(\exp
)}>M_{2}^{(\exp)}$, considered at the same body mass $M$. The finding
(\ref{CU-pred}) is in accord with the experimental observation [44]: muscles
contracting nearly isometrically (strut function) generate force more
economically than muscles involved in concentric work (via motor function).

\subsubsection{Muscle dynamics of mammalian legs and dragonfly wings}

Given that \emph{mammalian leg extensors} are active mostly during lengthening
[2], the brake primary function ($m=2$ in Table 2) could be assigned to leg
muscles specified by effective length $L_{leg}\varpropto M^{1/4}$
($\alpha_{leg}=0$ is adopted). In accord with Hill's second constraint,
underlaid by the proper frequency $T_{3m}^{-1}\varpropto L_{m}^{-2}$ (Table
4), the theory predicts $V_{leg}^{(\max)}\varpropto L_{leg}^{-1}\varpropto
M^{-1/4}$ that results in $1/T_{leg}^{(\max)}\varpropto L_{leg}^{-2}\varpropto
M^{-1/16}$. Similarly, for the wing-motor muscles in \emph{flying birds}
($m=1$ in Table 2) one should expect $V_{wing}^{(\max)}\varpropto
L_{wing}^{-1}\varpropto M^{-1/5}$, for contraction velocity, and
$1/T_{wing}^{(\max)}\varpropto M^{-1/25}$, for the frequency\ or,
alternatively, $1/T_{wing}^{(opt)}\varpropto M^{-1/5}$, in the
optimum-velocity regime (see Table 4). Hence, analytically revealed Hill's
constraint becomes observable via the empirical regression data by Medler
[43]: on the maximum-amplitude contraction velocities for the locomotor
muscles in leg of terrestrial animals, $V_{leg}^{(\exp)}\varpropto M^{-0.25}$,
and that for wings in flying birds, bats, and insects, $V_{wing}^{(\exp
)}\varpropto M^{-0.20}$. Moreover, the experimental data by Schilder and
Marden [45] of the wingbeat frequency $1/T_{wing}^{(\exp)}\varpropto
M_{m}^{-0.20}$ scaled by mass $M_{m}$ (and length $L_{0m}$) of the basalar
muscle in dragonflies indicate that the motor-type muscles ($L_{0m}\varpropto
M_{m}^{1/5}$, see analysis in Fig. 6) were studied self-consistently in the
optimum, steady-velocity motion regime.

.

\textbf{Place Fig. 6}

.

In the same optimum-velocity regime (Table 1), the maximum-amplitude
\emph{static force} $F_{stat}^{(\exp)}\cong\Delta F_{1m}^{(slow)}\varpropto
M_{m}^{2/3}$ and net\textrm{\ }\emph{lever-system force} $F_{ind}^{(\exp
)}\cong\Delta F_{1m}^{(fast)}\varpropto M_{m}$ reported by Schilder and Marden
[45] may be associated with the slow and fast activated fibres in the basalar
muscles tuned elastically to the linear regime through the dynamic PCSA
$A_{1m}^{(dyn)}\varpropto M_{m}^{2/3}$ and length $L_{1m}^{(dyn)}\varpropto
M_{m}^{1/3}$. The observed dynamic force output $F_{dyn}^{(\exp)}\varpropto
M_{m}^{0.83}$\ can be therefore suggested as the mixed-fibre force
$F_{dyn}^{(pred)}\cong\Delta F_{1m}^{(mix)}\varpropto M_{m}^{5/6}$ (Table 1),
\textit{i.e.} as a compromise of the forces $F_{stat}^{(\exp)}$ and
$F_{ind}^{(\exp)}$. These estimates challenge further analysis of the reported
dynamic forces.

\section{Conclusion}

A theoretical framework for mechanical characterization of the three transient
activated states of the striated muscles passing in force-length cycles
through the three distinct dynamic regimes is proposed. The explicit
analytical description of muscle locomotor functions and related mechanical
characteristics is provided on the basis of two concepts: (i) the preservation
of spindle-type shape in skeletal muscles and egg-type shape in cardiac
muscles related to the preservation of dynamic muscle volume and (ii) the
mechanical similarity between action and reaction forces emerging in
biomechanically equivalent states. Exploring known patterns of elastic forces
in continuum mechanics, the macroscopic study of the force production and its
functional and structural accommodation in the loaded muscle organ as a whole
provides the following major points.

1. It is demonstrated how the dynamic (frequency-velocity) constraints for
muscle contractions, first observed by Hill in hovering birds and then
revealed in locomotor muscles of running animals and flying birds, bats, and
insects, can be derived from the generic principle of mechanical (force and
velocity) similarity.

2. It is shown how relations in classical mechanics of solids can be explored
in soft tissues. The study is grounded by the active-force muscle stiffness
reliably derived in all muscle work loops nearby and below the
maximum-amplitude exerted forces. The muscle stiffness, underlaid by sarcomere
stiffness, is shown to be dependent on muscle geometry and dynamic functional
variable, underlaid by elastic moduli, which encompass all contractive
elements acting as an elastic continuum medium.

3. The theoretical prediction that the fast and slow muscles should generate
maximum forces linear, respectively, with the muscle volume and
cross-sectional area, regardless of muscle function and structure, is in part
validated by the direct empirical observation of maximal forces exerted by
animals and by the provided indirect observation of the adapted (primary)
muscle functions in legs of mammals and birds.

4. The macroscopic structures of locomotor skeletal muscles observable
directly by muscle allometry are found to be adapted to the maximum-force
state, following moderate-velocity dynamic regime, instead of the expected
optimum velocity regime.\ Such a bilinear-displacement muscle dynamics
involving both fast-twitch and slow-twitch powering muscle fibres sheds light
on the origin of allometric power laws and muscle specialization. The adapted
structures are examined via available empirical data: the legs are brakes in
mammals and springs in non-running birds, whereas the wings are motor-brake
engines in flying species. Suggested pump function for the cardiac muscles
needs further experimental tests.

5. The provided study of the muscle specialization in mammalian hindlimb
indicates that the properly tuned force production function is a dominated
factor in the accommodation of muscle structure. This finding also indicates
the predomination role of mechanical effects over biological adaptive
mechanisms assigned to the relatively small muscle-mass index. As the result,
a new investigation tool for indirect statistical observation of the
biomechanical adaptation of individual locomotor muscles is proposed through
the regression analysis of \textit{in vivo} muscle stresses in synergists
scaled across different-sized animals.

6. The assumption on that the muscle tuning muscle ability of animals can be
modeled by active elastic forces via non-linear muscle elastic moduli is
validated by the observation of the theoretical predictions for muscle
dynamics of legs and wings in running and flying specialists. Predictions are
made for the experimental modelling the primary and secondary function by
tuning the cycling muscle to the corresponding natural frequency and
controlling its efficiency.

7. The conservative character of architecture and related mechanical
characteristics of striated muscles suggests general trends following from
mechanical and shape constraints. The trends dictated by primary functions
explain, in particular, why the muscles having larger fibre and sarcomere
lengths and suited to efficient eccentric work, tend toward higher optimum
contraction velocities, but show lower maximum stiffness and mechanical energy cost.

8. As an intriguing outcome of the analysis of maximal contraction muscle
velocities and frequencies, the maximum-speed steady locomotion is revealed to
be \emph{controlled} by non-linear elasticity of slow-fibre muscles generating
moderated force. This finding deserves further evaluation in finite muscle
element analysis studying top speeds of living and extant animals.

.

\textbf{Acknowledgments}

I thank Andrew A. Biewener and James H. Marden for careful reading of the
draft of this paper and helpful critical comments. Rudolf J. Schilder and
James H. Marden are appreciated for giving datapoints in Fig. 6. The financial
support by CNPq is also acknowledged.

.

\textbf{Appendix A}. List of abbreviations\textbf{ }

.

PCSA - physiologic cross-sectional area

.

\textit{Mathematical signs and symbols }

$=$ - common equality sign

$\equiv$ - identity sign implying "by definition"

$\approx$ - approximate equality sign

$\sim$ - proportionality relation symbol omitting only numerical coefficients

$\cong$ - here used as similarity sign supporting only physical dimension units

$\propto$ - here used as scaling rule symbol not supporting dimension units

.

\textit{Physical and geometrical notations}

$\alpha_{m}$- muscle-mass allometric index

$\varepsilon_{m}^{(opt)}$- muscle strain in the optimum dynamic regime

$\eta_{m}$- muscle geometry parameter

$\rho_{tiss}$- tissue density

$\sigma_{tiss}^{(\max)}$- peak tissue stress

$\Delta L$ - length change

$\Delta F$ - force change

$\Delta t_{m}$- activation timing of muscle $m$

$A_{rm}$- cross-sectional area of muscle $m$ in passive ($r=0$) and active
($r\neq0$) states

$a$- scaling exponent for cross-sectional area

$D$ - diameter of ideal cylinder

$E_{rm}$ - active-muscle elastic modulus establishing the dynamic regime
$r\neq0$

$e$- strain scaling exponent

$\Delta F_{elast}^{(\max)}=\Delta F_{m}^{(\max)}$- maximum active elastic force

$F_{prod}^{(fast)}$- production force by fast muscle

$F_{motor}^{(conc)}$- elastic force adapted to concentric work in motor muscle

$F_{musc}^{(\max)}$- maximum force exerted by muscle

$K_{m}$- active muscle stiffness

$K_{s}$- sarcomere/cellular stiffness

$L$- length of an ideal cylinder

$L_{m}$ -\ variable muscle length in non-specified dynamics

$L_{rm}$ - dynamic muscle length in the regime $r$

$l$ - length exponent

$m$ - muscle in unspecified function

$M$ - body mass of animals

$M_{m}$ - muscle mass

$r$ - numerical parameter indicating transient dynamic states via
optimum-velocity ($r=1$), moderate-velocity ($r=2$), and high-velocity ($r=3$)
dynamic regimes, distinct of passive muscle state ($r=0$).

$T_{rm}$- period of cycling in the adapted regime $r$

$V_{rm}$- muscle contraction velocity in the dynamic regime $r$

$W$ - body weight

.

\textbf{Appendix B. }Scaling Muscle Functions

.

The \emph{motor function} is associated with the active\textrm{\ }%
force\textrm{\ }$F_{prod}^{(\max)}$ generated during muscle shortening at
moderate contraction velocity at the turning points $2$ in Figs. 1A, 1B, and
1C. In \emph{fast-fibre} muscles, the corresponding \emph{concentric force}%

\begin{equation}
F_{motor}^{(conc)}=F_{elast}^{(\max)}=\Delta F_{2m}^{(conc)}\sim
E_{2m}^{(fast)}A_{2m}^{3/2}L_{2m}^{-1}\text{ }\cong F_{prod}^{(fast)}
\label{F-bend}%
\end{equation}
is described by the known universal pattern of the maximal elastic forces [33]
equally applied to pure bending, pure torsion, or complex bending-torsion
loads subjected to long cylinder of length $L_{2m}$ and cross-sectional area
[26] $A_{2m}$. The exploration of Eq. (\ref{F-bend}) though Eqs.
(\ref{am-lm}), (\ref{Fm-elast}), (\ref{Ffaslo}), and (\ref{E2}) results in the
fast-muscle-force constraint $3a_{m}/2-l_{m}=1+\alpha_{m}$. It is remarkable
that the case of slow-fibre muscle, namely%
\begin{equation}
F_{motor}^{(conc)}=F_{elast}^{(\max)}=\Delta F_{2m}^{(conc)}\sim
E_{2m}^{(slow)}A_{2m}^{3/2}L_{2m}^{-1}\text{ }\cong F_{prod}^{(slow)}
\label{F-bend-slow}%
\end{equation}
results in the slow-muscle-force constraint $3a_{m}/2-2l_{m}=a_{m}$, which is
exactly the same as fast muscle, in view of \ function-independent Eq.
(\ref{am-lm-new}). Therefore, any muscle tuned to the motor locomotor function
should expose its \emph{dynamic structure} scaled by%
\begin{equation}
a_{motor}^{(conc)}=\frac{4}{5}(1+\alpha_{motor})\text{ , }l_{motor}%
^{(conc)}=\frac{1}{5}(1+\alpha_{motor})\text{, } \label{motor}%
\end{equation}
regardless of the fibre type content. This finding follows from both the
muscle force constraints solved with the help of the function-independent
muscle-shape constraint (\ref{am-lm-new}). Moreover, as shown in [26], the
principal component of the compressive stress $\sigma_{m}^{(conc)}$ specifying
Eq. (\ref{sigma-el}) may be caused by the peak transverse-tensile
\emph{strains}%
\begin{equation}
\varepsilon_{motor}^{(conc)}=\frac{\Delta D_{m}^{(\max)}}{L_{m}}\varpropto
M_{m}^{e_{m}}\text{, with }e_{m}=e_{motor}^{(conc)}=\frac{a_{m}}{2}%
-l_{m}\text{,} \label{etr-m}%
\end{equation}
where $\Delta D_{m}^{(\max)}\thicksim D_{m}\backsim A_{m}^{1/2}$ is transverse
muscle deformation.

Likewise, the maximum elastic \emph{eccentric force}
\begin{equation}
F_{brake}^{(eccen)}=\Delta F_{2m}^{(eccen)}\sim E_{2m}^{(eccen)}A_{2m}%
^{2}L_{2m}^{-2}\cong F_{prod}^{(fast)}\text{,} \label{Fm-stret}%
\end{equation}
associated with the \emph{brake muscle function} (Fig. 1B) provides the
maximum elastic stress%
\begin{equation}
\sigma_{brake}^{(\max)}=\frac{F_{brake}^{(eccen)}}{A_{m}}\varpropto
M_{m}^{s_{m}}\text{, with }s_{m}=s_{brake}^{(eccen)}=a_{m}-2l_{m}\text{,}
\label{p-crit}%
\end{equation}
following from Eqs. (\ref{sigma-el}) and (\ref{Fm-stret}). The unique solution
to both fast-muscle-force constraint, $2a_{m}-2l_{m}=1+\alpha_{m}$, and
slow-muscle-force constraint, $2a_{m}-3l_{m}=a_{m}$, is%
\begin{equation}
a_{brake}^{(eccen)}=\frac{3}{4}(1+\alpha_{brake})\text{ , }l_{brake}%
^{(eccen)}=\text{ }s_{brake}^{(eccen)}=\frac{1}{4}(1+\alpha_{brake})\text{.}
\label{brake}%
\end{equation}

The \emph{strut muscle function} treated as antagonistic to both motor and
brake functions drives nearly isometric contractions characteristic of small,
but non-zero length change ($\Delta L_{m}\ll L_{m}$) achieved near peak forces
(see Fig. 1C). This suggests the nearly \emph{isometric force}%
\begin{equation}
F_{strut}^{(isom)}=E_{2m}^{(isom)}\varepsilon_{2m}^{(isom)}A_{2m}\cong
F_{prod}^{(fast)}\text{, with }\varepsilon_{2m}^{(isom)}=\Delta L_{2m}%
^{(isom)}/L_{2m}\text{,} \label{Fm-strut}%
\end{equation}
in fast muscles. Again, one solves the \emph{muscle strut }constraints
$2a_{m}+l_{m}=1+\alpha_{m}$ and $2a_{m}+2l_{m}=a_{m}$ resulting in%
\begin{equation}
a_{strut}^{(isom)}=1+\alpha_{strut}\text{ and }l_{strut}^{(isom)}%
=s_{strut}^{(isom)}=0\text{, with }\Delta L_{2m}^{(isom)}\varpropto L_{2m}%
^{2}\text{,} \label{strut}%
\end{equation}
for any type of muscles.

A new antagonist (to strut muscle) tuned to the \emph{cardiac} type
contractions via active elastic force%
\begin{equation}
F_{pump}^{(card)}=\Delta F_{2m}^{(card)}\thicksim E_{2m}^{(card)}L_{2m}%
^{2}\cong F_{prod}^{(fast)} \label{Funk}%
\end{equation}
is associated with, say, \emph{pump function} providing the fast-muscle-force
constraint $2l_{m}=1+\alpha_{m}$. This yields
\begin{equation}
a_{pump}^{(card)}=l_{pump}^{(card)}=\text{ }s_{pump}^{(card)}=e_{pump}%
^{(card)}=\frac{1}{2}(1+\alpha_{pump})\text{,} \label{plung}%
\end{equation}
equally applied to slow-fibre muscles resulting in the slow-force constraint
$l_{m}=a_{m}$.

To complete the intrinsic-force description, the spring-type \emph{control
function }associated with the optimum-regime elastic force%
\begin{equation}
F_{contr}^{(sprin)}=F_{elast}^{(opt)}\varpropto E_{1m}^{(slow)}M_{m}%
^{2/3}\varpropto E_{1m}^{(slow)}A_{m}^{2/3}L_{m}^{2/3}\cong F_{prod}^{(slow)}
\label{F-contr}%
\end{equation}
in slow-fiber muscles results in%
\begin{equation}
a_{cont}^{(sprin)}=\frac{2}{3}(1+\alpha_{cont})\text{, }l_{cont}%
^{(sprin)}=\frac{1}{3}(1+\alpha_{cont})\text{, with }s_{cont}^{(sprin)}%
=e_{cont}^{(sprin)}=0\text{,} \label{spring}%
\end{equation}
that follows from the slow-force and fast-force constraints $2(a_{m}%
+l_{m})/3=a_{m}$\ and\ $(2a_{m}+5l_{m})/3=1+\alpha_{m}$ and therefore is valid
for any type of muscle tuned to velocity-optimum regime. All obtained
specific-function mechanical characteristics are summarized in Table 2.

.

\textbf{References}

.

[1] A.V. Hill, The dimensions of animals and their muscular dynamics, Science
Progr. 38 (1950) 209.

[2] S.L. Lindstedt, E.R. Trude, K. Paul , C.L. Paul, Do muscle function as
adaptable locomotor springs?, J. Exp. Biol. 205 (2002) 2221.

[3] Lichtwark GA, Wilson AM. Effects of series elasticity and activation
conditions on muscle power output and efficiency. J Exp Biol 208 (2005) 2845.

[4] Squire JM. Architecture and function in the muscle sarcomere. Curr Opin
Struct Biol 1997; 7: 247-55.

[5] Russell B, Motlagh D, Ashley WW. Form follows function: how muscle shape
is regulated by work. J Appl Physiol 2000; 88: 1127--32.

[6] Dickinson MH, Farley CT, Full JR, Koehl MAR, Kram R, Lehman S. How animals
move: an integrative view. Science 2000; 288: 100-6.

[7] Rome LC. Design and function of superfast muscles: New insights into the
physiology of skeletal muscle. Ann Rev Physiol 2006; 68: 193-221.

[8] Medler S, Hulme K. Frequency-dependent power output and skeletal muscle
design. Comp Biochem Physiol A 2009; doi:10.1016/j.cbpa.2008.11.021, in press.

[9] Jontes JD. Theories of muscle contraction. J Struct Biol 1995; 115: 119-43.

[10] Cole van den GK, Bogert AJ, Herzog W, Gerritsen KGM. Modelling of force
production in skeletal muscle undergoing stretch. J Biomech 1996; 29: 1091-104.

[11] Alexander RMcN. Optimum muscle design for oscillatory movements. J Theor
Biol 1997; 184: 253-9.

[12] Forcinito M, Epstein M, Herzog W.Can a rheological muscle model predict
force depression/enhancement? J Biomech 1998; 31: 1093-9.

[13] Kokshenev VB. A force-similarity model of the activated muscle is able to
predict primary locomotor functions. J Biomech 2008; 41: 912-5.

[14] Jenkyn TR, \ Koopman B, Huijing P, Lieber RL, Kaufman KR. Finite element
model of intramuscular pressure during isometric contraction of skeletal
muscle. Phys Med Biol 2002; 47: 4043--61.

[15] Marra SP, Ramesh KT, Douglas AS. Characterization and modeling of
compliant active materials. J Mech Phys Solids 2003; 51: 1723 -- 43.

[16] Skatulla S, Arockiarajan A, Sansour C. A nonlinear generalized continuum
approach for electro-elasticity including scale effects. J Mech Phys Solids
2009; 57: 137--60.

[17] Dumont ER, Grosse IR, Slater GJ. Requirements for comparing the
performance of finite element models of biological structures. J Theor Biol
2009; 256: 96--103.

[18] McMahon TA. Size and shape in biology. Science 1973; 179: 1201-4.

[19] McMahon TA. Using body size to understand the structural design of
animals: quadrupedal locomotion. J. Appl. Physiol. 39, 619-627.

[20] Kokshenev VB. Observation of mammalian similarity through allometric
scaling laws. Physica A 2003; 322: 491-505.

[21] Alexander RMcN. Allometry of the limbs of antelopes (Bovidae). J Zool
Lond 1977; 183: 125-46.

[22] Pollock CM, Shadwick RE. Allometry of muscle, tendon, and elastic
energy-storage capacity in mammals. Am J Physiol 1994; 266: R1022-31.

[23] Bennett MB. Allometry of the leg muscles in birds. J Zool Lond 1996; 238: 435-43.

[24] Biewener AA, Konieczynski DD, Baudinette RV, 1998. \textit{In vivo}
muscle force-length behavior during steady-speed hopping in tammar wallabies.
J Exp Biol 1998; 201: 1681-94.

[25] Biewener AA. Biomechanical consequences of scaling. J Exp Biol 2005; 208: 1665-76.

[26] Kokshenev VB. New insights into long-bone biomechanics: Are limb safety
factors invariable across mammalian species? J Biomech 2007; 40: 2911-18.

[27] Roberts TJ, Marsh RI, Weyand PG, Taylor CR. Muscular force in running
turkey: the economy of minimizing work. Science 1997; 275: 1113-5.

[28] Rome LC. Testing \ a muscle's design. Am Scientist 1997; 85: 356-8.

[29] Josephson RK. Dissecting muscle power output. J Exp Biol 1999; 202: 3369-75.

[30] Ahlborn BK, Blake RW, Megill WM. Frequency tuning in animal locomotion.
Zoology 2006; 109: 43--53.

[31] Forcinito M, Epstein M, Herzog W. Theoretical considerations on myofibril
stiffness. Biophys J 1997; 72: 1278-86.

[32] Robinson JM, Wang Y, Kerrick WGL, Kawai R, Cheung HC. Activation of
striated muscle: nearest-neighbor regulatory-unit and cross-bridge influence
on myofilament kinetics. J Mol Biol 2002; 322: 1065--88.

[33] Landau LD, \ Lifshitz EM. Theory of Elasticity. London: Pergamon Press; 1989.

[34] Kent GC. Comparative anatomy of the vertebrates, 7rd ed. Dubuque: Wm. C.
Brown Publishers; 1987.

[35] Rome LC, Runke RP, Alexander RMcN, Lutz G, Aldridge, H, Scott F,
Freadman, M. Why animals have different fibre types. Nature 1988; 335: 824--27.

[36] Farley CT, Glasheen J, McMahon TA. Running springs: speed and animal
size. J Exp Biol 1993; 185: 71-86.

[37] Bejan A, Marden JH. Unifying constructal theory for scale effects in
running, swimming and flying. J Exp Biol 2006; 209: 238-48.

[38] Rome LC, Sosnicki AA, Goble DO. Maximum velocity of shortening of three
fibre types from horse soleus muscle: implications for scaling with body size.
J Physiol 1990; 431: 173--85.

[39] Marden JH, Allen LR. Molecules, muscles, and machines: universal
performance characteristics of motors. Proc. Natl. Acad. Sci. USA 2002; 99: 4161-66.

[40] Daley MA, Biewener AA. Muscle force-length dynamics during level versus
incline locomotion: a comparison of in vivo performance of two guinea fowl
ankle extensors. J Exp Biol 2003; 206: 2941-58.

[41] Davis J, Kaufman KR, Lieber RL. Correlation between active and passive
isometric stress and intramuscular pressure in the isolated rabbit tibialis
anterior muscle. J Biomech 2003; 36: 505-12.

[42] Maloiy GMO, Alexander RMcN, Njau R, Jayes AS. Allometry of legs of
running birds. J Zool Lond 1979; 187: 161-67.

[43] Medler S. Comparative trends in shortening velocity and force production
in skeletal muscles. Am. J. Physiol. Regulatory Integrative Comp Physiol 2002;
283: R368--78.

[44] Demetrius L. The origin of allometric scaling laws in biology. J Theor
Biol 2006; 243:\textbf{\ }455-67.

[45] Schilder JR, Marden JH. A hierarchical analysis of the scaling of force
and power production by dragonfly flight motors. J Exp Biol 2004; 207: 767-76.

[46] J. Mendez, A. Keys, Density and composition of mammalian muscle,
Metabolism 9 (1960) 184.\newpage

\textbf{Figure Captions}

.

\textbf{Fig. 1}.{} The qualitative analysis of the \textit{in vivo} muscle
force-length data. The muscle \emph{motor function} is presented by
gastrocnemius powering during shortening in uphill running turkey (\emph{inset
A}, adapted from [27]). The lateral gastrocnemius and plantaris act as brake
(\emph{inset B}) and strut (\emph{inset C}) in hopping tammar wallabies [24].
The solid (and dashed) \emph{arrows} indicate rasing (and decreasing) of the
exerted force near its maximum magnitude $F_{\max}$. The regions of the linear
force-length domain are displayed by the force change $\Delta F_{1m}$ and
length change $\Delta L_{1m}$, both estimated from $L_{1m}$, and the starting
datapoint $F_{1m}$ of the force enhancement achieved at the optimum
contraction velocity $V_{opt}$. Similar to physical pendulum, the resting
length $L_{0m}$ is expected to be passed at near maximum velocities $V_{\max}$
and lower forces $F_{3m}$. The origin of intrinsic muscle forces (\emph{inset
D}): in both cases of the powering shortening (\emph{motor}) and lengthening
(\emph{brake}) muscles the resulted force $F_{\max}$ is a superposition of the
production force output $F_{prod}$ and reaction passive $F_{pass}$ and active
$F_{act}$ \emph{elastic} forces [see also text below Eq. (\ref{Fm-elast})].

.

\textbf{Fig. 2}. The indirect observation of the primary activity of mammalian
plantaris. The \emph{solid symbol} is the datapoint [22] presented in Table 5
and the bars indicate experimental error. The \emph{open symbols} are
theoretical estimates for stable dynamic structures established for the motor,
brake, strut, or control functions described in Table 2, with $\alpha
_{m}=\alpha_{0m}^{(est)}$ taken from Table 5.

.

\textbf{Fig. 3}. {}The observation of the primary mechanical function in some
isolated individual muscles in mammals. The analysis and notations correspond
to those in Fig. 2. The experimental (and theoretical) data for gastrocnemius,
DDF (deep digital flexor), and CDE (common digital extensor) are shown,
respectively, by the closed (and open) inverted triangles, regular triangles,
and circles. All the data are taken from Table 5.

.

\textbf{Fig. 4}. The\textbf{\ }qualitative study of the \textit{in vivo} data
on the peak stress in individual leg muscles of animals in strenuous activity.
The symbols employed above in Figs. 2\ and 3 are extended by the \textit{open
circles} (triceps) for the data on peak muscle stress taken from Table 1 in
[25], with the exclusion of the slow-mode data on cantering goat and trotting
cat. The data [41] on the activated isometric stress in isolated white rabbit
tibialis are added. The \emph{dashed line} shows the brake-functional stress
indicated by the stress scaling exponent $s=1/4$. The \emph{solid lines} are
drawn by $115\cdot M^{1/5}$, for the motor function, and by $215$ \emph{kPa},
for the strut and spring functions. All coefficients are adjusted by eye.

.

\textbf{Fig. 5}. The analysis of the primary mechanical functions for leg
muscles in running and non-running birds. The measured (and estimated) data
taken from Table 6 (and Table 2) for gastrocnemius, femorotibialis, and
digital flexors are shown by the closed (and open) inverted triangles,
circles, and regular triangles, respectively. The semi-open triangles are the
data by Bennett [23] for non-running birds.

.

\textbf{Fig. 6}. The qualitative scaling of the basalar structure to muscle
mass in male dragonflies (Odonata and Anisoptera, listed in Fig. 5 in [45].
The datapoints for muscle length $L_{0m}^{(\exp)}$ is a courtesy by the
authors. The estimated muscle cross-sectional area $A_{0m}^{(est)}$ is
obtained on the basis of Eq. (\ref{Ro-m}) taken with $\rho_{musc}^{(\exp)}=$
$1060$ $\emph{kg/m}^{3}$ [46]. The \emph{solid lines} are $L_{motor}%
=0.052\cdot M_{m}^{1/5}$ and $A_{motor}=0.018\cdot M_{m}^{4/5}$. \ The
\emph{dashed lines} indicated by the scaling exponents are drawn according to
muscle specialization shown in Table 2. All pre-exponential coefficients are
adjusted by eye.\pagebreak

\textbf{Tables}

.%

\begin{tabular}
[c]{|l|l|l|l|}\hline
Optimum muscle characteristics, (equations) & Fast fibres & Slow fibres &
Mixed fibres\\\hline\hline
Optimum length change, $\Delta L_{1m}$, (\ref{E1}) & $L_{m}$ & $L_{m}$ &
$L_{m}$\\\hline
Production/active-elastic force, $\Delta F_{1m}$, {\small (\ref{Fm-elast}),
(\ref{Felast}), (\ref{E1})} & $A_{m}L_{m}$ & $A_{m}$ & $A_{m}L_{m}^{1/2}%
$\\\hline
Optimum stiffness, $K_{1m}=E_{1m}A_{1m}/L_{1m}$, (\ref{Kopt}) & $A_{m}$ &
$A_{m}L_{m}^{-1}$ & $A_{m}L_{m}^{-1/2}$\\\hline
Optimum elastic\textrm{\ }stress, $\sigma_{1m}=\Delta F_{1m}/A_{1m}$,
(\ref{sigma-el}) & $L_{m}$ & $L_{m}^{0}$ & $L_{m}^{1/2}$\\\hline
Contraction frequency, $T_{1m}^{-1}\thicksim\sqrt{E_{1m}/\rho_{0m}}/L_{1m}$,
(\ref{Freq}) & $L_{m}^{-1/2}$ & $L_{m}^{-1}$ & $L_{m}^{-3/4}$\\\hline
Optimum velocity, $V_{1m}=V_{musc}^{(opt)}$, (\ref{Vcont}) & $L_{m}^{1/2}$ &
$L_{m}^{0}$ & $L_{m}^{1/4}$\\\hline
Optimum power, $P_{1m}=F_{1m}V_{1m}$ & $A_{m}L_{m}^{3/2}$ & $A_{m}$ &
$A_{m}L_{m}^{3/4}$\\\hline
\end{tabular}

\textbf{Table 1}. General mechanical characteristics of the striated muscles
tuned to linear-displacement dynamic regime scaled to dynamic fiber length
$L_{m}=L_{1m}$. The \emph{mixed-fibre} scaling dynamic exponents (shown in the
last column) are modeled by the common means for the fast-muscle and
slow-muscle exponents (established in the second and third columns),
\textit{i.e.} $F_{mix}\backsim\sqrt{F_{fast}F_{slow}}$; $A_{m}$ and $L_{m}$
are attributed to the stabilized \emph{dynamic} muscle geometry constrained by
muscle volume (\ref{Mmu}).

.

\begin{center}%
\begin{tabular}
[c]{|c|c|c|c|c|c|}\hline
${\small
\begin{tabular}
[c]{c}%
Locomotor pattern, regime\\
\ (equation)
\end{tabular}
}$ & ${\small
\begin{tabular}
[c]{c}%
Motor, r=2\\
\ (\ref{motor})
\end{tabular}
}$ &
\begin{tabular}
[c]{c}%
{\small Brake, r=2}\\
{\small (\ref{brake})}%
\end{tabular}
{\small \ } &
\begin{tabular}
[c]{c}%
{\small Strut, r=2}\\
{\small \ (\ref{strut})}%
\end{tabular}
{\small \ } &
\begin{tabular}
[c]{c}%
{\small Control, r=1}\\
{\small \ (\ref{spring})}%
\end{tabular}
{\small \ } &
\begin{tabular}
[c]{c}%
{\small Pump, r=2}\\
{\small \ (\ref{plung})}%
\end{tabular}
{\small \ }\\\hline\hline
${\small
\begin{tabular}
[c]{c}%
Force pattern, muscle\\
\ (equation)
\end{tabular}
}$ &
\begin{tabular}
[c]{c}%
$F_{{\small motor}}^{{\small (conc)}}$,{\small \ m=1}\\
{\small \ (\ref{F-bend})}%
\end{tabular}
&
\begin{tabular}
[c]{c}%
$F_{{\small brake}}^{{\small (eccen)}}$,{\small \ m=2}\\
\multicolumn{1}{l}{{\small (\ref{Fm-stret})}}%
\end{tabular}
{\small \ } &
\begin{tabular}
[c]{c}%
$F_{{\small strut}}^{{\small (isom)}}$, {\small m=3}\\
{\tiny \ }{\small (\ref{Fm-strut})}%
\end{tabular}
{\small \ } &
\begin{tabular}
[c]{c}%
$F_{{\small contr}}^{{\small (sprin)}}$, {\small m=4}\\
\multicolumn{1}{l}{{\tiny \ }{\small (\ref{F-contr})}}%
\end{tabular}
{\small \ } &
\begin{tabular}
[c]{c}%
$F_{{\small pump}}^{{\small (card)}}$, {\small m=5}\\
\multicolumn{1}{l}{{\tiny \ }{\small (\ref{Funk})}}%
\end{tabular}
{\small \ }\\\hline
\multicolumn{1}{|l|}{{\small Maximum force output, (\ref{Ffaslo})}} &
$1+\alpha_{1}$ & $1+\alpha_{2}$ & $1+\alpha_{3}$ & $\frac{2}{3}(1+\alpha_{4})$
& $1+\alpha_{5}$\\\hline
\multicolumn{1}{|l|}{{\small Muscle fibre length, (\ref{am-lm})}} & $\frac
{1}{5}(1+\alpha_{1})$ & $\frac{1}{4}(1+\alpha_{2})$ & $0$ & $\frac{1}%
{3}(1+\alpha_{4})$ & $\frac{1}{2}(1+\alpha_{5})$\\\hline
\multicolumn{1}{|l|}{{\small Cross-sectional area, (\ref{am-lm})}} & $\frac
{4}{5}(1+\alpha_{1})$ & $\frac{3}{4}(1+\alpha_{2})$ & $1+\alpha_{3}$ &
$\frac{2}{3}(1+\alpha_{4})$ & $\frac{1}{2}(1+\alpha_{5})$\\\hline
\multicolumn{1}{|l|}{{\small Structure parameter, }$\eta_{m}$=$a_{m}l_{m}%
^{-1}${\small \ }} & $4$ & $3$ & $\ \infty$ & $2$ & $1$\\\hline
\multicolumn{1}{|l|}{{\small Length change*, (\ref{Fmax})}} & $\frac{2}%
{5}(1+\alpha_{1})$ & $\frac{1}{2}(1+\alpha_{2})$ & $0$ & $\frac{1}{3}%
(1+\alpha_{4})$ & $1+\alpha_{5}$\\\hline
\multicolumn{1}{|l|}{{\small Maximum stress/strain*, (\ref{Sm})}} & $\frac
{1}{5}(1+\alpha_{1})$ & $\frac{1}{4}(1+\alpha_{2})$ & $0$ & $0$ & $\frac{1}%
{2}(1+\alpha_{5})$\\\hline
\multicolumn{1}{|l|}{{\small Maximum stiffness*, (\ref{Kopt})}} & $\frac{3}%
{5}(1+\alpha_{1})$ & $\frac{1}{2}(1+\alpha_{2})$ & $1+\alpha_{3}$ & $\frac
{1}{3}(1+\alpha_{4})$ & $0$\\\hline
\multicolumn{1}{|l|}{{\small Natural} {\small frequency*, (\ref{Freq})}} &
$-\frac{1}{5}(1+\alpha_{1})$ & $-\frac{1}{4}(1+\alpha_{2})$ & $0$ & $-\frac
{1}{3}(1+\alpha_{4})$ & $-\frac{1}{2}(1+\alpha_{5})$\\\hline
\multicolumn{1}{|l|}{{\small Energy change*, (\ref{U-m})}} & $\frac{7}%
{5}(1+\alpha_{1})$ & $\frac{3}{2}(1+\alpha_{2})$ & $1+\alpha_{3}$ &
$1+\alpha_{4}$ & $2(1+\alpha_{5})$\\\hline
\multicolumn{1}{|l|}{{\small Moderate velocity* (\ref{Vcont}) }} & $0$ & $0$ &
$0$ & $0$ & $0$\\\hline
\end{tabular}

\end{center}

\textbf{Table 2}. The locomotor functions and their mechanical characteristics
scaled to dynamic structures. The all-type powering individual muscles
$m=1,2,3,$ and $5$ are tuned to the maximum-force bilinear dynamic regime
$r=2$ [described in Eq. (\ref{E2})] and muscles $m=4$ act in the linear regime
$r=1$ [Eq. (\ref{E1})]. *The data shown for fast muscles. The allometric
exponents are related to animal's body mass via Eq. (\ref{am-lm})\textbf{.}

.%

\begin{tabular}
[c]{|l|l||l|c||l|c||l|c||l|c|}\hline
{\small Dyn. regimes} & {\small Force} & {\small Funct.} & {\small Structure}
& {\small Funct.} & {\small Structure} & {\small Funct.} & {\small Structure}
& {\small Funct.} & {\small Structure}\\\hline
$r=1,2,3$ & ${\small F}_{{\small prod}}^{{\small \max}}${\small \ }%
${\small \varpropto}$ & ${\small F}_{{\small motor}}^{{\small conc}%
}{\small \varpropto}$ &
\begin{tabular}
[c]{|l|l|}\hline
${\small a}_{m}$ & ${\small l}_{m}$\\\hline
\end{tabular}
& ${\small F}_{{\small brake}}^{{\small eccen}}{\small \varpropto}$ &
\begin{tabular}
[c]{|l|l|}\hline
${\small a}_{m}$ & ${\small l}_{m}$\\\hline
\end{tabular}
& ${\small F}_{{\small strut}}^{{\small isom}}{\small \varpropto}$ &
\begin{tabular}
[c]{|l|l|}\hline
${\small a}_{m}$ & ${\small l}_{m}$\\\hline
\end{tabular}
& ${\small F}_{{\small plun}}^{{\small card}}{\small \varpropto}$ &
\begin{tabular}
[c]{|l|l|}\hline
${\small a}_{m}$ & ${\small l}_{m}$\\\hline
\end{tabular}
\\\hline\hline
${\small E}_{1m}^{(slow)}{\small \varpropto L}_{m}^{{\small 0}}$ &
${\small A}_{m}$ & ${\small A}_{m}^{\frac{3}{2}}{\small L}_{{\small m}%
}^{{\small -}1}$ &
\begin{tabular}
[c]{l|l}%
$\frac{{\small 2}}{3}$ & $\frac{{\small 1}}{3}$%
\end{tabular}
& ${\small A}_{m}^{2}{\small L}_{m}^{-2}$ &
\begin{tabular}
[c]{l|l}%
$\frac{{\small 2}}{3}$ & $\frac{{\small 1}}{3}$%
\end{tabular}
& ${\small A}_{m}$ &
\begin{tabular}
[c]{l|l}%
${\small nc}$ & ${\small nc}$%
\end{tabular}
& ${\small L}_{m}^{2}$ &
\begin{tabular}
[c]{l|l}%
$\frac{{\small 2}}{3}$ & $\frac{{\small 1}}{3}$%
\end{tabular}
\\\hline
${\small E}_{1m}^{(fast)}{\small \varpropto L}_{m}$ & ${\small A}_{{\small m}%
}{\small L}_{{\small m}}$ & ${\small A}_{m}^{\frac{3}{2}}$ &
\begin{tabular}
[c]{l|l}%
$\frac{{\small 2}}{3}$ & $\frac{{\small 1}}{3}$%
\end{tabular}
& ${\small A}_{m}^{2}{\small L}_{m}^{-1}$ &
\begin{tabular}
[c]{l|l}%
$\frac{{\small 2}}{3}$ & $\frac{{\small 1}}{3}$%
\end{tabular}
& ${\small A}_{m}{\small L}_{m}$ &
\begin{tabular}
[c]{l|l}%
${\small nc}$ & ${\small nc}$%
\end{tabular}
& ${\small L}_{m}^{3}$ &
\begin{tabular}
[c]{l|l}%
$\frac{{\small 2}}{3}$ & $\frac{{\small 1}}{3}$%
\end{tabular}
\\\hline
{\small \ }${\small E}_{2m}^{(slow)}{\small \varpropto L}_{{\small m}%
}^{{\small -1}}$ & ${\small A}_{m}$ & ${\small A}_{m}^{\frac{3}{2}}%
{\small L}_{{\small m}}^{{\small -2}}$ &
\begin{tabular}
[c]{l|l}%
$\frac{\mathbf{4}}{\mathbf{5}}$ & $\frac{\mathbf{1}}{\mathbf{5}}$%
\end{tabular}
& ${\small A}_{m}^{2}{\small L}_{m}^{-3}$ &
\begin{tabular}
[c]{l|l}%
$\frac{\mathbf{3}}{\mathbf{4}}$ & $\frac{\mathbf{1}}{\mathbf{4}}$%
\end{tabular}
& ${\small L}_{m}^{-1}{\small A}_{m}$ &
\begin{tabular}
[c]{l|l}%
$\mathbf{1}$ & $\mathbf{0}$%
\end{tabular}
& ${\small L}_{m}^{1}$ &
\begin{tabular}
[c]{l|l}%
$\frac{\mathbf{1}}{\mathbf{2}}$ & $\frac{\mathbf{1}}{\mathbf{2}}$%
\end{tabular}
\\\hline
{\small \ }${\small E}_{2m}^{(fast)}{\small \varpropto}L_{m}^{0}$ &
${\small A}_{{\small m}}{\small L}_{{\small m}}$ & ${\small A}_{{\small 0}%
}^{\frac{3}{2}}{\small L}_{{\small 0}}^{{\small -1}}$ &
\begin{tabular}
[c]{l|l}%
$\frac{\mathbf{4}}{\mathbf{5}}$ & $\frac{\mathbf{1}}{\mathbf{5}}$%
\end{tabular}
& ${\small A}_{0}^{2}{\small L}_{0}^{-2}$ &
\begin{tabular}
[c]{l|l}%
$\frac{\mathbf{3}}{\mathbf{4}}$ & $\frac{\mathbf{1}}{\mathbf{4}}$%
\end{tabular}
& ${\small A}_{0}$ &
\begin{tabular}
[c]{l|l}%
$\mathbf{1}$ & $\mathbf{0}$%
\end{tabular}
& ${\small L}_{0}^{2}$ &
\begin{tabular}
[c]{l|l}%
$\frac{\mathbf{1}}{\mathbf{2}}$ & $\frac{\mathbf{1}}{\mathbf{2}}$%
\end{tabular}
\\\hline
${\small E}_{3m}^{(slow)}{\small \varpropto L}_{m}^{{\small -2}}$ &
${\small A}_{m}$ & ${\small A}_{m}^{\frac{3}{2}}{\small L}_{{\small m}%
}^{{\small -3}}$ &
\begin{tabular}
[c]{l|l}%
$\frac{{\small 6}}{7}$ & $\frac{{\small 1}}{7}$%
\end{tabular}
& ${\small A}_{m}^{2}{\small L}_{m}^{-4}$ &
\begin{tabular}
[c]{l|l}%
$\frac{{\small 4}}{5}$ & $\frac{{\small 1}}{5}$%
\end{tabular}
& ${\small L}_{m}^{-2}{\small A}_{m}$ &
\begin{tabular}
[c]{l|l}%
${\small 1}$ & ${\small 0}$%
\end{tabular}
& ${\small L}_{m}^{0}$ &
\begin{tabular}
[c]{l|l}%
${\small 0}$ & ${\small 1}$%
\end{tabular}
\\\hline
${\small E}_{3m}^{(fast)}{\small \varpropto L}_{m}^{{\small -1}}$ &
${\small A}_{{\small m}}{\small L}_{{\small m}}$ & ${\small A}_{m}^{\frac
{3}{2}}{\small L}_{{\small m}}^{{\small -2}}$ &
\begin{tabular}
[c]{l|l}%
$\frac{{\small 6}}{7}$ & $\frac{{\small 1}}{7}$%
\end{tabular}
& ${\small A}_{m}^{2}{\small L}_{m}^{-3}$ &
\begin{tabular}
[c]{l|l}%
$\frac{{\small 4}}{5}$ & $\frac{{\small 1}}{5}$%
\end{tabular}
& ${\small A}_{m}{\small L}_{m}^{-1}$ &
\begin{tabular}
[c]{l|l}%
${\small 1}$ & ${\small 0}$%
\end{tabular}
& ${\small L}_{m}^{1}$ &
\begin{tabular}
[c]{l|l}%
${\small 0}$ & ${\small 1}$%
\end{tabular}
\\\hline
\end{tabular}

\textbf{Table} \textbf{3}. Locomotor functions predicted by dynamic structured
for slow and fast striated muscles tuned to distinct dynamic regimes. The
primary functions ($r=2$) are shown by bold type. The analysis of functional
muscle structures made in terms of elastic-force patterns: the active-muscle
optimum-velocity ($r=1$), moderate-velocity ($r=2$), and high-velocity ($r=3$)
dynamic regimes are described in the first column via the muscle elastic
moduli $E_{rm}$ [Eqs. (\ref{E1}), (\ref{E2}), and (\ref{E3})] and specified by
slow and fast force output [Eq. (\ref{Ffaslo})], shown in the second column.
The third and next odd columns show the elastic force functional scaling in
concentric, eccentric, isometric, and pump contractions. The corresponding
solutions to scaling equations underlaid by the force similarity principle
(\ref{Fm-elast}) are shown for simplicity with $\alpha_{rm}=0$, in the forth
and next even columns. \emph{Notation:} $nc$ indicates non-conclusive solution.

.%

\begin{tabular}
[c]{|l|ll|ll|ll|}\hline
Dynamic regimes & Optimum, & $r=1$ & Moderate, & $r=2$ & Maximum, &
$r=3$\\\hline
Muscle type & \multicolumn{1}{|c}{slow} & \multicolumn{1}{|c|}{fast} &
\multicolumn{1}{|c}{slow} & \multicolumn{1}{|c|}{fast} &
\multicolumn{1}{|c}{slow} & \multicolumn{1}{|c|}{fast}\\\hline\hline
Natural frequency, Eq. (\ref{Freq}) & \multicolumn{1}{|c}{$L_{m}^{-1}$} &
\multicolumn{1}{|c|}{$L_{m}^{-1/2}$} & \multicolumn{1}{|c}{$L_{m}^{-3/2}$} &
\multicolumn{1}{|c|}{$L_{m}^{-1}$} & \multicolumn{1}{|c}{$L_{m}^{-2}$} &
\multicolumn{1}{|c|}{$L_{m}^{-3/2}$}\\\hline
Contraction velocity, Eq. (\ref{Vcont}) & \multicolumn{1}{|c}{$L_{m}^{0}$} &
\multicolumn{1}{|c|}{$L_{m}^{1/2}$} & \multicolumn{1}{|c}{$L_{m}^{-1/2}$} &
\multicolumn{1}{|c|}{$L_{m}^{0}$} & \multicolumn{1}{|c}{$L_{m}^{-1}$} &
\multicolumn{1}{|c|}{$L_{m}^{-1/2}$}\\\hline
\end{tabular}

\textbf{Table} \textbf{4}. Dynamic characterization of the red (slow) and
white (fast) striated muscles in the optimum-, moderate-, and maximum-velocity
dynamic regimes $r=1,2$, and $3$ described in Table 3.

.%

\begin{tabular}
[c]{|l|c|c|c|c|c|c|c|l|}\hline
{\small Individual mammalian muscles} & $a_{0m}^{(\exp)}$ & $l_{0m}^{(\exp)}$
& ${\small \alpha}_{0m}^{(\exp)}$ & ${\small \eta}_{{\small 0}m}$ &
${\small \alpha}_{0m}^{(est)}$ & $\mathit{a}_{m}$ & $\mathit{l}_{m}$ &
{\small Prim. functions}\\\hline\hline
{\small Gastrocnemius (and soleus)} & ${\small 0.77\pm}.{\small 02}$ &
${\small 0.21\pm.02}$ & {\small -}${\small 0.03}$ & ${\small 3.7}$ &
{\small -}${\small 0.02}$ & ${\small 0.78}$ & ${\small 0.20}$ & {\small motor,
}${\small m=1}$\\\hline
{\small Deep digital flexor (DDF)}$^{\ast)}$ & ${\small 0.85\pm.03}$ &
${\small 0.18\pm.02}$ & ${\small \ 0.03}$ & ${\small 4.7}$ & ${\small \ 0.03}
$ & ${\small 0.82}$ & ${\small 0.21}$ & {\small motor, }${\small m=1}$\\\hline
{\small Comm. digit. extensor (CDE)} & ${\small 0.69\pm.04}$ &
${\small 0.24\pm.02}$ & {\small -}${\small 0.07}$ & ${\small 2.9}$ &
{\small -}${\small 0.07}$ & ${\small 0.70}$ & ${\small 0.23}$ & {\small brake,
}${\small m=2}$\\\hline
{\small Plantaris (SDF)} & ${\small 0.91\pm.04}$ & ${\small 0.05\pm.04}$ &
{\small -}${\small 0.03}$ & ${\small 18}$ & {\small -}${\small 0.04}$ &
${\small 0.96}$ & ${\small 0.00}$ & {\small strut, }${\small m=3}%
$\\\hline\hline
{\small Ankle-joint muscle group} & ${\small 0.81\pm.03}$ & ${\small 0.17\pm
.03}$ & {\small -}${\small 0.03}$ & ${\small 4.8}$ & {\small -}${\small 0.03}$
& ${\small 0.78}$ & ${\small 0.19}$ & {\small motor, }${\small g=1}$\\\hline
\end{tabular}

\textbf{Table 5}. The analysis of the allometric data by Pollock and Shadwick
[22] provided on the basis of Eq. (\ref{am-lm-new}) and Table 2. The shown
statistical error is approximated by the symmetrized $95\%$ confidence
interval. The methodology of the analysis is illustrated in Fig. 2. The
primary functions found in Figs. 2 and 3 are described following Table 2, with
$\alpha_{m}=\alpha_{0m}^{(est)}$. The overall muscle group ($g=1$) is
determined as the standard mean over all muscles. $^{\ast)}$DDF includes
individual flexor hallucis and flexor digitorum longus; SDF means superficial
digital flexor.

.%

\begin{tabular}
[c]{|l|c|c|c|c|c|c|l|}\hline
Running birds & $\ a_{0m}^{(\exp)}$ & $\alpha_{0m}^{(\exp)}$ & $l_{0m}%
^{(est)}$ & $a_{0m}^{(\exp)}/l_{0m}^{(est)}$ & $a_{2m}$ & $l_{2m}$ & Primary
function/force\\\hline\hline
Gastrocnemius & $0.81\pm0.14$ & $\ 0.14$ & $0.33$ & $2.5$ & $0.85$ & $0.29$ &
brake/eccentric\\\hline
Digital flexors (DF) & $0.76\pm0.22$ & $-0.03$ & $0.21$ & $3.6$ & $0.78$ &
$0.19$ & motor/concentric\\\hline
Femorotibialis & $0.80\pm0.12$ & $-0.02$ & $0.18$ & $4.4$ & $0.78$ & $0.20$ &
motor/concentric\\\hline\hline
Overall group & $0.79\pm0.16$ & $\ 0.03$ & $0.24$ & $3.3$ & $0.77$ & $0.26$ &
brake/eccentric\\\hline
\end{tabular}

\textbf{Table 6. }The analysis of the allometric data by Maloiy \textit{et
al.} [42]. The shown large error is due to relatively wide confidence limits.
The mean exponents $l_{0m}^{(est)}$ are estimated via Eq. (\ref{am-lm-new}).
The overall muscle group is determined as the standard mean over all muscles.
The indicated primary functions and active elastic forces are described by the
evaluated dynamic-structure exponents $a_{2m}$ and $l_{2m}$ found as most
close to the experimental resting-volume data on $a_{0m}^{(\exp)}$ and
$l_{0m}^{(\exp)}$ and therefore assigned to regime $r=2$ (Table 2).

\end{document}